\documentclass[useAMS,usenatbib,times]{mn2e}
\usepackage{epsfig,times}

\title[Weak lensing with GEMS]{Cosmological weak lensing with the HST GEMS survey}
\author[Heymans et al.]{Catherine Heymans$^{1}$\thanks{heymans@mpia.de},
Michael L. Brown$^{2}$, Marco Barden$^1$, John
A.R. Caldwell$^{3,4}$,\newauthor  
Knud Jahnke$^5$, Chien Y. Peng$^{3,6}$,
Hans-Walter Rix$^1$,  Andy Taylor$^2$, \newauthor Steven
V. W. Beckwith$^{3,7}$, 
Eric F. Bell$^1$, Andrea Borch$^1$, Boris H\"au\ss ler$^1$, Shardha
Jogee$^{3,8}$, \newauthor Daniel H. McIntosh$^9$, Klaus 
Meisenheimer$^1$, 
 Sebastian F. S\'anchez$^{4,10}$, Rachel Somerville$^3$, \newauthor
Lutz Wisotzki$^4$ \& 
Christian Wolf$^{11}$.\\
$^1$Max-Planck-Institut f\"{u}r Astronomie, K\"{o}nigstuhl, D-69117,
Heidelberg, Germany\\
$^2$Institute for Astronomy, University of Edinburgh, Royal Observatory,
Blackford Hill, Edinburgh, EH9 3HJ, UK \\
$^3$Space Telescope Science Institute, 3700 San Martin Drive, Baltimore, MD
    21218, USA.\\
$^4$University of Texas, McDonald Observatory
Fort Davis, TX 79734, USA.\\
$^5$Astrophysikalisches Insitut Potsdam, An der Stenwarte 16, 14482 Potsdam,
    Germany.\\
$^6$Steward Observatory, University of Arizona, 933 N. Cherry Ave., Tuscon, AZ
    85721, USA.\\
$^7$Department of Physics and Astronomy, The Johns Hopkins University, 3400
    North Charles Street, Baltimore, MD 21218, USA.\\
$^8$Department of Astronomy, University of Texas at Austin, 1 University
    Station, C1400 Austin, TX 78712-0259, USA.\\
$^9$Department of Astronomy, University of Massachusetts, 710 North Pleasant
    Street, Amherst, MA 01003, USA.\\
$^{10}$Centro Hispano Aleman de Calar Alto, C/Jesus Durban Remon 2-2, E-04004
    Almeria, Spain.\\
$^{11}$Department of Astrophysics, Denys Wilkinson Building, University of
    Oxford, Keble Road, Oxford, OX1 3RH, UK.
}

\newcommand{\be}{\begin{equation}}  \newcommand{\ee}{\end{equation}}
\newcommand{\bes}{\begin{equation*}}  \newcommand{\ees}{\end{equation*}}
  \newcommand{\ba}{\begin{eqnarray}}
\newcommand{\ea}{\end{eqnarray}}  
\newcommand{\nn}{\nonumber\\}  
\newcommand{\bx}{{\bf x}}
\newcommand{\myplus}{+}
\newcommand{\myminus}{-}

\newcommand{\bm}[1]{\mbox{\boldmath{$#1$}}}

\newcommand{\lgl}{\langle}
\newcommand{\rgl}{\rangle}
\def\gs{\mathrel{\raise1.16pt\hbox{$>$}\kern-7.0pt %
\lower3.06pt\hbox{{$\scriptstyle \sim$}}}}         %
\def\ls{\mathrel{\raise1.16pt\hbox{$<$}\kern-7.0pt %
\lower3.06pt\hbox{{$\scriptstyle \sim$}}}}         %

\begin{document}


\pagerange{\pageref{firstpage}--\pageref{lastpage}} \pubyear{2004}
\maketitle
\label{firstpage}

\begin{abstract}
We present our cosmic shear analysis of GEMS, one of the largest wide-field
surveys ever undertaken by the Hubble Space Telescope.  Imaged with the
Advanced Camera for Surveys (ACS), GEMS spans 795 square
arcmin in the Chandra Deep Field South.  
We detect weak lensing by large-scale structure in high resolution 
F606W GEMS data from $\sim 60$ resolved galaxies per square 
arcminute.  We measure the two-point shear correlation function,
the top-hat shear variance and the shear power spectrum, performing an E/B mode
decomposition for each statistic.  We show that we are not limited by
systematic errors and use our results to place joint constraints on the matter
density parameter $\Omega_m$ and the
amplitude of the matter power spectrum $\sigma_8$.  We find
$\sigma_8(\Omega_m/0.3)^{0.65}=0.68 \pm 0.13$ where the $1\sigma$ 
error includes both our
uncertainty on the median redshift of the survey and sampling variance.

Removing image and point
spread function (PSF) distortions are crucial to all weak lensing analyses. 
We therefore include a thorough discussion on the degree of ACS PSF
distortion and anisotropy which we characterise directly from GEMS data.
Consecutively imaged over 20
days, GEMS data also allows us to investigate PSF instability
over time.  We find that, even in the relatively short GEMS observing
period, the ACS PSF ellipticity varies at the level of a few percent which we
account for with a semi-time dependent PSF model.  Our correction for the
temporal and spatial variability of the PSF is shown to be successful through
a series of diagnostic tests.

\end{abstract}


\section{Introduction}

Weak gravitational lensing is a unique probe of the dark matter distribution 
at redshifts $z<1$ where, within the currently favoured $\Lambda$CDM
cosmological paradigm, 
dark energy begins to play an important role in the evolution and growth of the
power spectrum of matter fluctuations.  It therefore not only
has the power to constrain fundamental cosmological parameters such as 
the matter density parameter $\Omega_m$ and the
amplitude of the matter power spectrum $\sigma_8$
\citep{Maoli,RRG01,vWb01,HYG02,BMRE,Jarvis,MLB02,Hamana,Massey,RhodesSTIS,vWb04},
but also has the potential to test and constrain quintessence models 
parameterised by 
the equation of state of the dark energy ${\rm w}(z)$ 
\citep{RefSNAP03,Jarvis05}.
Cosmological parameter constraints from weak lensing analysis
are fully complementary to those from cosmic microwave
background (CMB) experiments 
as parameter degeneracies are almost orthogonal
in many cases \citep{MLB02,Contaldi,Tereno}.  Alone, measurements of the CMB
anisotropy at  
$z\sim1000$ are unable to constrain ${\rm w}(z)$ but combined with future wide-field 
weak lensing surveys, potentially focused on selected galaxy clusters
\citep{JainTay}, 
the goal of determining ${\rm w}(z)$ will certainly become attainable. 

Lensing by large-scale structure distorts images of background galaxies, 
inducing weak correlations in the observed 
ellipticities of galaxies, termed `cosmic
shear'.   The amplitude and angular dependence of these
correlations are related to the non-linear matter power spectrum $P_\delta(\ell)$ and the
geometry of the Universe.  If we first
assume that the minute weak lensing shear distortions can be
measured in the absence of any systematic errors, arising for example
from telescope and
detector based distortions, the error on any weak lensing analysis then has 
four
main sources: shot noise from the intrinsic ellipticity distribution of
galaxies, sampling (cosmic) variance, uncertainty in the 
galaxy redshift distribution and uncertainty in intrinsic 
correlations that may exist 
between nearby galaxy pairs.  Shot noise can be minimised by
surveying large areas of sky and/or by increasing the number density of
resolved galaxies in the weak lensing analysis.  Sampling variance 
can be minimised by imaging many different lines of sight.
With spectroscopic or 
photometric redshift information the galaxy redshift distribution can be
accurately estimated \citep{MLB02,vWb04}.  Redshift information also allows
for the removal of intrinsic correlations
\citep{KingSch02,HH03,HBH04}.  This effect has however been
shown to be a small contaminant to the weak lensing signal measured
from deep surveys (median redshift $z_m
\sim 1.0$), at the level of less than a few percent of the 
shear correlation signal \citep{HBH04}.
The next generation of weak lensing surveys aim to obtain precise results,
minimising these sources of error by
surveying of the order of a hundred square degrees with accompanying
photometric redshift 
information. These ground-based surveys are however subject  
to atmospheric seeing which erases the weak lensing shear information
from all 
galaxies smaller than the size of the seeing disk.  This, in effect, limits the
maximum depth of ground-based weak lensing surveys and hence the sensitivity,
leading to  proposals for future deep wide-field space-based 
observations \citep{RhodesSNAP03}.  With space-based data the number density
of resolved galaxies, in comparison to ground-based studies, is
multiplied two-fold and more.  Resolving images of more distant galaxies
will permit high resolution maps of the dark matter distribution
\citep{MaseySNAP03} and will yield significantly higher signal-to-noise
constraints on  
cosmological parameters in comparison to constraints obtained from
the same survey area of a seeing limited 
ground-based survey (Brown et al. in prep).

With the installation of the Advanced Camera for Surveys
(ACS) on the Hubble Space Telescope (HST), relatively wide-field space-based
weak lensing studies 
are feasible and in this paper we present the detection of weak
gravitational lensing by large-scale structure in GEMS; the Galaxy
Evolution from Morphology and Spectral energy distributions survey (see
\citealt{RixGEMS} for an overview).
Spanning $795$ square arcmin, GEMS 
is currently the largest space-based colour mosaic. 
With a high number density of resolved sources and spectroscopic and/or
photometric redshifts for $\sim 8000$ of our sources from the COMBO-17 survey
\citep{C17} and the VVDS \citep{VVDS}, we can beat down shot noise and obtain a
good estimate of the galaxy redshift distribution.

The error sources discussed above have not included systematic errors, but in
reality for 
all weak lensing surveys the accuracy of any analysis depends critically on
the correction for instrumental distortions, which are orders 
of magnitude larger than the 
underlying gravitational shear distortions that we wish to detect.
The strongest distortion results from a convolution
of the image with the point spread function (PSF) of the telescope and
camera. For the detection of cosmic shear 
at the level of only a few percent, the PSF distortion needs
to be understood and controlled to an accuracy of better than one percent.
For space-based weak lensing studies 
\citep{HFKS98,RRG00,RRG01,RRG02,STISsch,Casertano,RhodesSTIS,Miralles}, the limited
number of stars in each camera field of view provides insufficient 
coverage to map the PSF anisotropy.  This has
made it necessary to assume long-term
stability in the PSF pattern, often using PSF models derived from
archived images of globular clusters, with low order focus corrections,
or from Tiny Tim \citep{TinyTim} 
thereby critically limiting the accuracy of the 
correction for PSF distortions.  
In contrast to previous space-based lensing studies we are not
forced to assume long-term PSF stability as the GEMS PSF can be 
characterised directly from the wide-area data where all but three out of the
sixty-three ACS images were observed in the space of 20 days.  Hence we need
only assume short term PSF 
stability, which we test in Section~\ref{sec:PSF_stability}.  A detailed
investigation into the GEMS ACS PSF will be presented in Jahnke et al. (in
prep) with a view to many different astronomical applications. 

This paper is organised as follows. In Section~\ref{sec:method} we summarise
the basic theory that underpins cosmic shear studies and review the
particulars of our analysis.  We discuss, in Section~\ref{sec:data}, the
observations and data reduction, paying specific attention to potential sources
of image distortions.  We focus on characterising and correcting for the
temporally and spatially varying dominant PSF distortion in
Section~\ref{sec:PSF} and compare, in 
Section~\ref{sec:compVz}, galaxy shear
measured from the two GEMS passbands; F606W and F850LP. In
Section~\ref{redshift_dist} we estimate the redshift distribution of the GEMS
survey from COMBO-17 and VVDS redshift catalogues.  Our statistical analysis 
is detailed in Section~\ref{sec:analysis_cs} where we measure the shear
correlation function, the top-hat shear
variance statistic and the shear power
spectrum performing many diagnostic tests for
systematic errors.  These results are used to place joint constraints on  
$\Omega_m$ and $\sigma_8$ in Section~\ref{sec:cosmoparam}.  We conclude in
Section~\ref{sec:conc} with a comparison to results from other space-based and
ground-based cosmic shear analyses.

\section{Method}
\label{sec:method}

The theory of weak gravitational lensing by large structure, detailed in
\citet{Bible}, directly relates the
non-linear matter power spectrum $P_{\delta}$ to the observable weak lensing 
(complex) shear field $\gamma = \gamma_1 + i\gamma_2$ characterised through
the shear (convergence) power spectrum $P_\kappa$, 
\begin{equation} 
P_\kappa(\ell) = \frac{9 H_0^4 \Omega_m^2}{4c^4} \int_0^{w_H} dw \, 
\frac{g^2(w)}{a^2(w)} \, P_\delta \left( \frac{\ell}{f_K(w)},w \right),
\label{eqn:Pkappa} 
\end{equation}
where $w$ is the comoving radial distance, $a(w)$ is the dimensionless scale
factor, $H_0$ is the Hubble parameter, 
$\Omega_m$ the matter density parameter and $g(w)$ is a  
weighting function locating the lensed sources, 
\begin{equation} 
g(w) = \int_w^{w_H}\, dw'\ \phi(w') 
\frac{f_K(w'-w)}{f_K(w')}, 
\label{eqn:W} 
\end{equation} 
where $\phi\left[w(z)\right]dw$ is the observed number of galaxies in $dw$ and
$w_H$ is the horizon distance \citep{Sch98}. Note in this paper we
will assume a flat Universe with zero curvature where the comoving
angular diameter distance $f_K(w) = w$. 

In this paper we measure directly the shear power spectrum
$P_\kappa$, the 
shear correlation function $\langle
\gamma(\bm{\theta})\gamma(\bm{\theta} + \Delta \theta) \rangle$ which we
split into a tangential component
\be
\langle \gamma_{t} \gamma_{t} \rangle_\theta =
\frac{1}{4\pi}\int d\ell \,\ell \,P_\kappa(\ell) \, [ J_0(\ell\theta) + J_4(\ell\theta)],
\label{eqn:gtgt}
\ee 
and a rotated component
\be
\langle \gamma_{r} \gamma_{r} \rangle_\theta =
\frac{1}{4\pi}\int d\ell \,\ell \,P_\kappa(\ell) \, [ J_0(\ell\theta) - J_4(\ell\theta)],
\label{eqn:grgr}
\ee
and the
top-hat shear variance $\langle |\gamma|^2 \rangle_\theta$ measured in a circle
of angular radius $\theta$  
\be
\langle \gamma^2 \rangle_\theta = 
\frac{1}{2\pi\theta^2}\int \frac{d\ell}{\ell} \,P_\kappa(\ell) \, [ J_1(\ell\theta)]^2 \,.
\label{eqn:shearvar}
\ee
Note $J_i$ is the $i^{\rm th}$ order Bessel function of the first kind.
 
In order to exploit the straightforward physics of weak lensing, one requires
an estimate of the gravitational shear experienced by each galaxy.
\citet{KSB}, \citet{LK97} and \citet{HFKS98} (KSB+) prescribe a method to
invert the effects of 
the PSF smearing and shearing, recovering an unbiased shear
estimator uncontaminated by the systematic distortion of the PSF. Objects
are parameterised according to their weighted quadrupole moments
\be
Q_{ij} = \frac{\int \, d^2\theta \, W(\bm{\theta}) \,
  I(\bm{\theta}) \, \theta_i
  \theta_j} {\int d^2\theta \, W(\bm{\theta}) \,I(\bm{\theta}) },
\label{eqn:quadmom}
\ee
where $I$ is the surface brightness of the object, $\theta$ is the angular
distance from the object centre and $W$ is a Gaussian weight function of scale
length $r_g$, where $r_g$ is some measurement of galaxy size, for
example the half light radius.  For a
perfect ellipse, the weighted quadrupole moments are related to the
weighted ellipticity parameters $\varepsilon_\alpha$ by
\be
\left(
\begin{array}{c}
\varepsilon_1 \\
\varepsilon_2
\end{array}
\left)
= \frac{1}{Q_{11} + Q_{22}}
\right(
\begin{array}{c}
Q_{11} - Q_{22} \\
2Q_{12}
\end{array}
\right).
\label{eqn:ellipquad}
\ee
\citet{KSB} show that if the PSF
distortion can be described as a small but highly anisotropic
distortion convolved with a large circularly symmetric seeing disk,
then the ellipticity of a PSF corrected galaxy is given by
\be
\varepsilon^{\rm cor}_{\alpha} = \varepsilon^{\rm obs}_{\alpha} - P^{\rm
  sm}_{\alpha\beta}p_{\beta}, 
\label{eqn:ecor}
\ee
where $p$ is a vector that measures the PSF anisotropy, and $P^{\rm sm}$ is
the smear polarisability tensor given in \citet{HFKS98}.
$p(\bm{\theta})$ 
can be estimated from images of stellar objects at position
$\bm{\theta}$ by noting that a 
star, denoted throughout this paper with $^*$, imaged
in the absence of PSF distortions has zero ellipticity:  
$\varepsilon^{*\, {\rm cor}}_{\alpha} = 0$. Hence,  
\be
p_{\mu} = \left(P^{\rm sm *} \right)_{\mu \alpha}^{-1}\,
\varepsilon^{*{\rm obs}}_{\alpha} .
\label{eqn:pmu}
\ee
For space-based imaging, where PSFs deviate strongly from a Gaussian, this PSF
correction is mathematically poorly defined 
\citep{K00} such that it is important to calculate the PSF correction
vector $p$ not only as a function of galaxy position $\bm{\theta}$,
but also as a function of galaxy size $r_g$ \citep{HFKS98}.  This
rather unsatisfactory situation has prompted the development of alternative
methods \citep{RRG00,K00,Bernstein,shapelets,Masseyshapelets} but for the
purpose of this 
paper we will focus on the most commonly used KSB+ technique, deferring our
analysis with different techniques to a future paper. 

The isotropic effect of the PSF is to convolve galaxy images with a circular
kernel.  This makes objects appear rounder,
erasing shear information for galaxies smaller than the kernel size, which have
to be removed from the galaxy shear sample.  For the larger galaxies, this
resolution effect can be accounted for
by applying the pre-seeing shear polarisability tensor correction $P^\gamma$,
as proposed by \citet{LK97}, such that
\be
\varepsilon^{\rm cor}_{\alpha} = \varepsilon^{s}_{\alpha} + P^\gamma_{\alpha
  \beta}\gamma_{\beta}.
\label{eqn:eseeing}
\ee
where $\varepsilon^{s}$ is the true source ellipticity and $\gamma$ is the
pre-seeing gravitational shear.
\citet{LK97} show that
\be
P^\gamma_{\alpha \beta} = P^{\rm sh}_{\alpha \beta} - P^{\rm
  sm}_{\alpha \mu} \left(P^{\rm sm *} \right)_{\mu \delta}^{-1} P^{\rm
  sh *}_{\delta \beta},
\label{eqn:Pgamma}
\ee
where $P^{\rm sh}$ is the shear polarisability tensor given in \citet{HFKS98}
and $P^{\rm sm *}$ and $P^{\rm sh *}$ are the stellar smear and shear
polarisability tensors respectively.  This relation is only strictly true
when all values are measured from the
PSF de-convolved image which is difficult to create in practice.   
$P^\gamma$ is therefore calculated from the PSF
distorted images which produces very noisy measurements.  

Combining the PSF correction, equation (\ref{eqn:ecor}), and the $P^\gamma$
seeing correction, 
the final KSB+ shear estimator $\hat{\gamma}$ is given by
\be
\hat{\gamma}_{\alpha} = \left(P^\gamma \right)^{-1}_{\alpha
  \beta} \left[\varepsilon^{\rm obs}_{\beta} - P^{\rm
  sm}_{\beta\mu}p_{\mu}\right].
\label{eqn:shearest}
\ee
When averaging over many galaxies, assuming a random distribution of intrinsic
galaxy ellipticities, $\langle
\varepsilon^{s} \rangle = 0$, and hence $\langle \hat{\gamma} \rangle =
\gamma$, providing a good estimate for the gravitational shear. 

\section{The GEMS data}
\label{sec:data}
The GEMS survey \citep{RixGEMS} 
spans an area of $\sim 28' \times 28'$ centred on the 
Chandra Deep Field South (CDFS), combining 125 orbits of ACS/HST time with
supplementary data from the GOODS project
\citep{GOODS}.  78 ACS tiles have been imaged in 
F606W and 77 ACS tiles in F850LP, where the point source $5\sigma$
detection limits reach ${\rm m}_{606}= 28.3$ and ${\rm m}_{850} = 27.1$.  In
this 
section we review the data set, discuss the potential for bias within the
source catalogues and highlight possible sources of
image distortion, in addition to the strong anisotropic PSF distortion which
is characterised and corrected for in Section~\ref{sec:PSF}.   
A detailed account of the full
GEMS data reduction method will be presented by Caldwell et al. (in
prep). 

The ACS wide-field camera has a field-of-view $\sim 3.4 \times 3.4$ arcmin
comprising two $4096 \times 2048$ CCD chips of pixel scale $0.05$ arcsec 
\citep{ACSWFC}.
GEMS observes, in sequence, three separate exposures per ACS tile dithered by
$\sim 3$ arcsec, where the observation strategy has been designed in such a
way so as to bridge the inter-chip gap and 
provide sub-pixel oversampling of pixel scale $0.03$ arcsec in the final
co-added 
image ($\sim 7000 \times 7000$ pixels).  GOODS have employed a 
different observing strategy, using only two separate dithered exposures.  
In order to optimise the survey for
Supernova searches \citep{GOODSSN1a} the GOODS area is re-imaged in 5 different
epochs, but to obtain similar depths to the GEMS data and minimise the effects
of PSF time variation, we co-add only the exposures from
the first epoch of observations. This however leaves us with a
slightly shallower central region in the F606W GEMS mosaic which can
be seen from the median magnitude of each data set; ${\rm m}_{606}({\rm GEMS})
= 25.6$, ${\rm m}_{606}({\rm GOODS}) = 25.1$.  
We note that the 2 exposure GOODS dithering pattern will
result in a poorer cosmic ray rejection in the GOODS area of the GEMS mosaic,
and will impact somewhat on the PSF. 

Images from the ACS suffer from strong geometric distortions as a result of
the off-axis location of the camera within HST.  In addition, the HST optical
assembly and the ACS mirrors also induce distortions.  \citet{Geodistort}
accurately calibrate and model this distortion from dithered images of star
clusters.  With this model all tiles are drizzled onto a celestial pixel
grid using a version of the multidrizzle software \citep{multi-drizzle},
where the astrometry of each GEMS
tile is tied to the overall catalogue from the ground-based COMBO-17 $R$ band
image \citep{C17}. 
Averaging our final PSF corrected shear catalogues as a function of ACS chip
position, we find no evidence for any residual geometric distortions
remaining in our final multi-drizzled images.  

A second order geometric distortion arises from the effect of velocity
aberration \citep{velab}.  The HST guides on nearby stars peripheral to 
the ACS field of view, making small
corrections to keep the primary target on a fixed pixel.  It cannot however
correct for the isotropic plate-scale breathing which is of the order
$\sim 0.001\%$ over an orbit.
These small changes in pixel scale are allowed for and
corrected by our version of multidrizzle, provided the individual exposures
are on the same 
scale. Variation of scale during each observation would 
result in a slight blurring of the co-added 
images that increases radially from the centre. 
GEMS ACS tiles, observed over 1/7 of the orbit, suffer from pixel scale
variation that is, at maximum, a difference of $0.004$ arcsec corner to
corner.  Our results later show no significant variation in average galaxy
shear as a function of chip position and we hence conclude that
this effect is not significant
within our measurement accuracy. In
future data reductions of GEMS we will include a velocity aberration
correction to the pixel scale 
using an updated version of multi-drizzle.
   
We use the {\it SExtractor} software \citep{SExt} to detect sources on both
the F606W and F850LP imaging data, with the two-step approach described in
\citet{RixGEMS}.  In short, this method of combining two {\it SExtractor}
catalogues, one measured with high signal-to-noise detection thresholding,
and one measured with low 
signal-to-noise detection thresholding, allows us to find the best compromise
between detecting faint galaxies without deblending bright nearby galaxies
into many different components.  We also use {\it SExtractor} to determine the
weak variation of the sky background across the tile, which is then subtracted
for the 
following KSB+ analysis. We define galaxy size $r_g$ as the half light radius
measured by {\it SExtractor} ({\tt flux\_radius})\footnote{{\it SExtractor}
  analysis of image simulations have shown that the
  {\tt flux\_radius} does not accurately 
measure the true half-light radius (H\"au\ss ler in prep.).  A good 
  relationship between {\tt flux\_radius} and the input
  half light radius is seen but
  this relationship differs between disk
  and bulge dominated galaxies.  The value that we
  choose for $r_g$ does not change the results as the final $P^\gamma$
  correction, equation (\ref{eqn:Pgamma}), includes a correction for the $r_g$
  Gaussian weighting.  The choice of $r_g$ does however
  affect the noise on each measurement and so we 
  choose $r_g = f \times ${\tt flux\_radius} where $f$ is chosen to minimise
  the noise in the measured galaxy ellipticities.  We find $f=1$. Note that
  alternative ways of defining $r_g$ using {\it SExtractor} ellipticity
  measures (see for example \citealt{RRG00}) have been found to yield
  slightly noisier results with the GEMS data.}
and calculate 
weighted ellipticity parameters $\varepsilon_i$ and the 
shear and smear polarisability tensors; $P^{\rm sh}$ and $P^{\rm sm}$  for
each object in the {\it SExtractor} catalogue. 

The accuracy of the centroid determined by
{\it SExtractor} is directly linked to the accuracy of each KSB+ galaxy shear
measurement.  In the presence of non-isotropic centroiding 
errors there is also the potential for centroid bias
\citep{K00,Bernstein} which can arise if, for example, the errors  
in the $x$ direction exceed those in the $y$ direction producing a
tendency to bias towards an average galaxy shear in the $x$ direction.  The
GEMS galaxies have been modelled using the two-dimensional galaxy profile model
fitting code {\it GALFIT} \citep{GALFIT} which
finds the best fit PSF convolved  S\'ersic profiles to each galaxy, 
allowing the centroid to be a free parameter in the fit (see \citealt{Barden}
and H\"au\ss ler et al. in prep for details).  We can therefore test if we are
subject to centroid bias by 
comparing {\it GALFIT} and {\it SExtractor} centroids.  The average pixel
offset is consistent with zero in the $x$ direction, $\Delta x = -0.001 \pm
0.02$ and very close to zero in the $y$ direction $\Delta y = 0.03 \pm
0.02$.  Calculating the ellipticity of a mock circular Gaussian galaxy
$N=10^4$ times, assuming Gaussian distributed centroid errors with mean and
width as estimated by {\it GALFIT}, we find that 
for our smallest galaxies, centroid errors induce a systematic centroid bias 
$\left[e_1 = (-2.754 \pm 0.001) \times 10^{-4}, e_2 = (-7.14 \pm 0.02) \times
10^{-6}\right]$, which is negligible 
compared to our current measurement accuracy.
For larger galaxies centroid bias decreases.
Note that the GEMS {\it GALFIT} galaxy profile parameters 
cannot currently be used for weak
lensing studies as the PSF has been derived from co-added stellar images and
therefore does not allow for the spatial variation of the distortion.
With a spatially varying PSF model {\it GALFIT}
could be used for measuring galaxy shear, although this would be a 
time consuming process.

When compiling source catalogues one should consider selection bias 
where any preference to select galaxies oriented in
the same direction as the PSF \citep{K00} and galaxies that are
anti-correlated with the gravitational shear (and as a result appear more
circular) \citep{Hirata}, would bias the mean
ellipticity of the population.  Through simulations of artificial disc galaxy
light profiles, convolved with the ACS PSF (see H\"au\ss ler et al. in prep
for details), 
we see no significant selection bias when we introduce a ${\rm SNR}>15$
selection criteria (defining SNR = {\tt flux} / {\tt flux\_error}),
i.e within the noise of the sample,  
there is neither a preference for selecting
faint galaxies oriented with the PSF, nor a preference for selecting more
circular faint objects.  

To remove erroneous detections along the 
chip boundaries, diffraction spikes from stars,
satellite trails and reflection ghosts, 
each image catalogue is masked by hand 
using the method described in
\citet{odtdata}. Using weight maps to define the best regions
in each tile, we combine the masked catalogues from each ACS tile ensuring
that in the overlapping regions of neighbouring tiles, only the data from the
better tile is included.   
Note that the objects in overlapping regions are used
for consistency checks to test the accuracy of the galaxy shear measurement.

The charge transfer efficiency (CTE) of space-based instruments provides
another source of image distortion.  Objects with low signal-to-noise in low
sky background images tend to
bleed in the read-out direction of the CCD camera, causing an 
elongation of the objects that is correlated with the read-out direction and
the distance from the read-out amplifier.  Over time the CTE degrades,
increasing the magnitude of this effect \citep{Miralles,RhodesSTIS}.  
The read-out-amplifiers for the ACS
lie at each corner of the camera, with the read-out direction along the $y$
axis of the CCD  
and we find no correlation between the average PSF corrected galaxy shear
along the $y$ axis, and the galaxy distance from the
read-out-amplifiers. Hence, even though the CTE of the ACS 
wide-field camera has been shown to degrade with time \citep{ACSCTE}, 
we find no signature for CTE in our data, potentially a result of the 
data being observed relatively soon after the ACS installation. 

\section{Characterisation of, and correction for, the anisotropic ACS PSF}
\label{sec:PSF}
In this section we give a thorough account of the techniques which we have used
to characterise and correct for the anisotropic ACS PSF.  To date there has not
been such a large set of HST data imaged in a short time frame 
which can allow for a rigorous semi-time dependent PSF analysis. 
The PSF is characterised through images of non-saturated
stellar objects that are selected
through their locus in the size-magnitude plane. 
Stars can be easily identified, extending from bright
magnitudes to faint magnitudes into the main 
distribution of galaxies, remaining at one characteristic size.  
We use both the full width half maximum ({\tt FWHM}) and the half light radius
({\tt flux\_radius}) measured by {\it 
  SExtractor} as a definition of size to select the stellar objects, where
stellar candidates must lie along the stellar locus in both the {\tt
  FWHM}-magnitude plane and the {\tt flux\_radius}-magnitude plane.
This method selects $\sim 1300$ stellar objects in the F850LP images and
$\sim 900$ stellar objects in the F606W images.  Note that the loss
of stellar objects in the F606W arises from the increased number of
saturated stars and the use of more conservative cuts at fainter magnitudes to
avoid confusion with small faint galaxies.

\begin{figure} 
\centerline{\epsfig{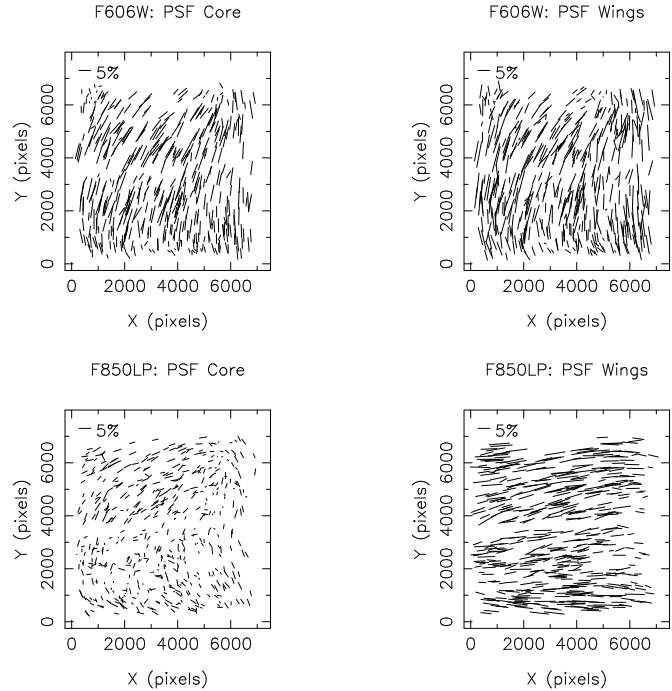}} 
\caption{The anisotropic ACS PSF measured from stellar sources in F606W images
  (upper panels) and F850LP  images (lower panels).  The stellar 
  ellipticity $\varepsilon^*$ plotted has been measured using two
  differently scaled weight functions $W(r_g)$; left panels $r_g = 2.5$ pixels
  (PSF 
  core distortion), right panels  $r_g = 7$ pixels (PSF wing distortion).  The
  5\% bar in the upper left corner of each panel shows the scale, which is the
  same for each panel.} 
\label{fig:Vz_core_wings} 
\end{figure}

For each star we measure the weighted 
stellar ellipticity parameters $\varepsilon^*_\alpha$ and the 
stellar smear polarisability tensor $P^{\rm sm *}$ using Gaussian
weights $W(r_g)$
with different smoothing scales; $r_g$ ranging from $1.9$ pixels,
the minimum stellar size measured by {\it SExtractor}, to $10$ pixels.  
We limit the maximum smoothing scale to avoid excessive inclusion of 
light from neighbouring objects which quickly introduces noise
into the stellar shape measurement.
Figure~\ref{fig:Vz_core_wings} shows the variation in the stellar
ellipticity parameters
$\varepsilon^*_{\mu}$ across the ACS field of view, measured using all the GEMS
F606W data, (upper panels) and all the GEMS F850LP data (lower
panels). The horizontal spacing at $y \sim 3200$ pixels results from the chip
boundaries where shape 
estimates become unreliable. For Figure~\ref{fig:Vz_core_wings},
$\varepsilon^*_{\alpha}$ has been measured using two
differently scaled weight functions $W(r_g)$, the first looking at the core PSF
distortion with $r_g = 2.5$ pixels 
(left panels), and the second looking at the PSF distortion averaged over the
main extent of the star, with $r_g = 7$ pixels (right panels).  This figure
shows that the PSF distortion is clearly anisotropic and
varies with scale size, and with filter.  The F850LP PSF has a strong
horizontal diffraction spike, which dominates the average PSF distortion 
on large scales resulting in a strong positive $\varepsilon_1$ component 
across the full field of view.   This diffraction spike  
may therefore account for the claim by 
\citet{Park} that the ACS PSF is fairly isotropic.

In order to accurately
characterise the anisotropy of the PSF across the field of view of each ACS
tile in the survey we wish to
maximise the surface density of stellar objects as a function of $(x,y)$
position.  This however necessitates some assumptions about the PSF stability
over time as the selected 
stellar number density corresponds to only $\sim 16$ stars
per ACS tile in F850LP and $\sim 11$ stars per ACS tile in F606W.
Figure~\ref{fig:Vz_core_wings} shows smooth variation in the PSF as a
function of chip position indicating that any variation of the PSF in the 20
day duration of the GEMS observations is small.  We split our stellar
sample into stars imaged by GEMS and stars imaged by GOODS, as each data set
derives from co-added exposures with different dithering patterns which 
impact on the PSF. The first epoch of GOODS observations 
spanned 5 days, and all but 3/63 GEMS tiles 
were observed in the space of 20 days. We reject from our analysis the 3
GEMS tiles which were taken out of sequence and split our GEMS sample into 2
data sets assuming PSF stability on the scale of 10 days.  We will
quantify the validity of this assumption in Section~\ref{sec:PSF_stability}.
Our PSF model is thus semi-time dependent as
we use three time-dependent PSF models derived from fits 
to three different data sets selected by observation date. 

\begin{figure} 
\centerline{\epsfig{file=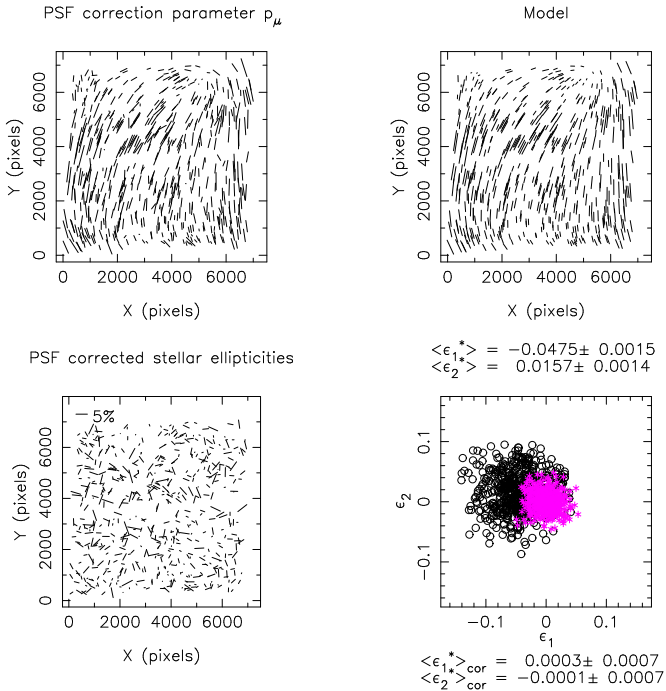,width=8.75cm,angle=0,clip=}} 
\caption{The upper left panel shows the 
variation in the measured PSF correction vector
$p$ across the ACS field of view,
measured using all the GEMS F606W data.  This data is modelled with a
semi-time dependent
two-dimensional second order polynomial shown in the upper right panel. 
The PSF corrected stellar ellipticities $\varepsilon^{\rm
  cor*}_{\mu}$, lower left panel, display random orientation.  The 
 5\% bar in the upper left corner shows the scale.  Note that 
$p$ and $\varepsilon$  are not directly equivalent and are 
thus plotted on different scales.  The
ellipticity distribution of stars before 
(circles) and after (dots) PSF correction are also shown (lower right).} 
\label{fig:PSF} 
\end{figure}

To model the anisotropy of the PSF across the field of view, we fit a 
two-dimensional second order polynomial to the PSF correction vector
$p$ equation (\ref{eqn:pmu}), modelling each CCD chip and data set separately.
Before 
fitting we remove outliers with a $3\sigma$ deviation from $\bar{p}_{\mu}$, 
and then iterate twice during the fit,
removing outliers with a $3\sigma$ deviation in their PSF corrected ellipticity
from $\bar{\varepsilon}^{\rm cor*}_{\mu}$.  Figure~\ref{fig:PSF} shows the
variation in the measured PSF correction vector
$p$ across the ACS field of view,
measured using all the GEMS F606W data (upper left). The $p$ values
calculated 
from our semi-time-dependent polynomial
models (upper right), the corrected stellar ellipticities $\varepsilon^{\rm
  cor*}_{\mu}$ (lower left, note
that $p$ and $\varepsilon$ are plotted on different scales), and the
ellipticity distribution of stars before 
and after the PSF correction (lower right) are also shown.  
For this figure we have used a
smoothing scale of $r_g = 5.9$ pixels, which is the median galaxy size in our
catalogue.  In this case we find the mean stellar ellipticity
before and after correction to be
\ba 
\bar{\varepsilon}_1^* = 0.0475 \pm 0.015 &
\bar{\varepsilon}_2^* = 0.0157 \pm 0.014 & {\rm (before),} \nn
\bar{\varepsilon}_1^* = 0.0003 \pm 0.0007 &
\bar{\varepsilon}_2^* = -0.0001 \pm 0.0007 & {\rm (after).} 
\label{eqn:bfaft}
\ea
This demonstrates that the PSF correction significantly reduces the mean
stellar ellipticity such 
that it is consistent with zero, and that the dispersion also decreases by a
factor $\sim 2$.
Note that the success of this correction does lessen somewhat
with increasing $r_g$, 
as the noise in the measurement of $p$ grows, but as the number of
galaxies to which these high $r_g$ corrections apply decreases in turn, this
effect is not problematic. 

\begin{figure} 
\centerline{\epsfig{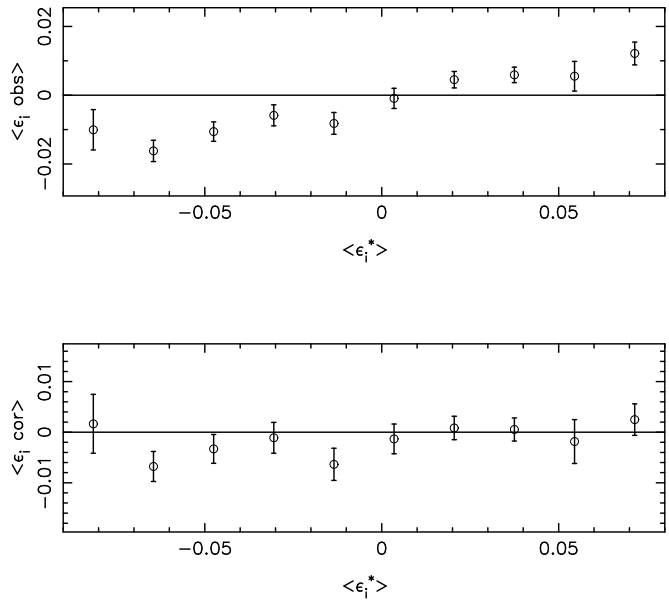}} 
\caption{The upper panel shows the correlation between 
 the mean observed galaxy ellipticity  
$\langle \varepsilon^{\rm   obs}_i \rangle$ averaged in square cells of side
  20 arcseconds, and the cell stellar ellipticity $\langle
 \varepsilon^{*}_i \rangle$ measured with $r_g = 5.9$ pixels.  Correction for
 the anisotropic PSF removes this correlation which can be seen in the lower
 panel where the mean PSF corrected galaxy ellipticity $\langle
 \varepsilon^{\rm   cor}_i \rangle$ is shown as a function
 of stellar ellipticity. Note that the upper and lower panels are plotted on
 different scales.}  
\label{fig:box} 
\end{figure}

With the PSF models we correct our galaxy catalogue for PSF distortions using
equation (\ref{eqn:ecor}). 
To test for residual PSF related systematic distortions 
we search for correlations
between PSF corrected galaxy ellipticity $\varepsilon^{\rm
  cor}_i $ and stellar ellipticity $\varepsilon_i^*$. 
Our first simple test splits the survey into square cells of side 20
arcseconds in the ACS $(x,y)$ plane, calculating the cell averaged PSF
corrected galaxy ellipticity $\langle \varepsilon^{\rm 
  cor}_i \rangle$ and the cell averaged uncorrected stellar ellipticity
$\langle \varepsilon_i^* \rangle$ determined at
$r_g = 5.9$, the median galaxy size in the survey. 
Figure~\ref{fig:box} (lower panel) shows the resulting mean 
$\Sigma \langle \varepsilon^{\rm   cor}_i \rangle 
/ N_{\rm cells}$ as a function of cell stellar ellipticity $\langle
\varepsilon_i^* \rangle$, where for comparison we also show the average
ellipticity of galaxies, which have not been PSF corrected, as a
function of stellar ellipticity (upper panel).
The correlation found with the uncorrected galaxy catalogue is not seen
in the PSF corrected galaxy ellipticities, indicating the success of the PSF
correction. This promising result will be tested more rigorously in
Section~\ref{sec:PSFtests}.

\subsection{The temporal stability of the PSF}
\label{sec:PSF_stability}

\begin{figure} 
\centerline{\epsfig{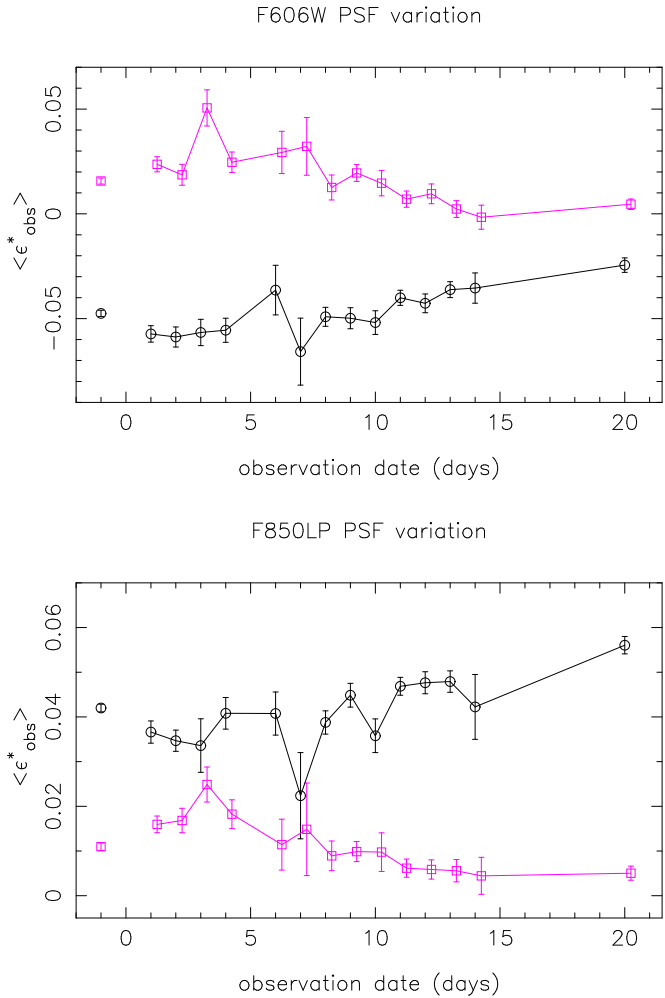}} 
\caption{The stellar ellipticity parameters $\varepsilon^*_1$ (circles) and
  $\varepsilon^*_2$ (squares) averaged across the full ACS field of
  view as a function of observation date.  The stellar 
  ellipticity averaged over the whole data set is plotted as an 
  isolated point on the far left of the plot.
  The upper panel shows the variation measured from F606W data and the lower
  panel shows the variation measured from F850LP showing that both filters
  follow the same trends in the $\varepsilon^*_1$ and $\varepsilon^*_2$
  components. The errors plotted are the errors on the mean stellar
  ellipticity.  For this figure we have used a smoothing
scale of $r_g = 5.9$ pixels to calculate the stellar
ellipticities, but we see PSF variation over the
full range of smoothing scales that we use in this analysis.    
} 
\label{fig:VzPSFvariation} 
\end{figure}

PSF time variation in space-based instruments is known to result from telescope
`breathing', as the HST goes into and out of sunlight in its 90 minute orbit,
and from a slow change in focus which is periodically corrected for
\citep{RRG00}. Variation in the PSF as measured from reduced
images can also be caused by slight differences in the data reduction method
but the consistent GEMS observation and reduction strategy minimises this
effect.  With our large set of HST data we are able 
to test the stability of the ACS PSF by looking for changes in the
average stellar ellipticity as a function of time and time dependent
changes in the anisotropy of the PSF distortion. 
Figure~\ref{fig:VzPSFvariation} shows the variation in the average F606W and
F850LP stellar 
ellipticity parameters as a function of observation date\footnote{Our
  consistent GEMS reduction strategy minimises the chances that the PSF
  variation we see results from data manipulation but the reader should note
  that the observation strategy of GEMS does correlate observation date and
  declination.} where for each ACS image the F850LP and F606W data were taken
in succession. This figure reveals
a clear trend in both filters with $\varepsilon^*_1$ (circles)
increasing and $\varepsilon^*_2$ (squares) decreasing by a few percent during
the observation period, a variation that is of the order of the signal we wish
to detect. 
This measurement of ACS PSF temporal instability is
in agreement with \citet{JeeACS} who show that their PSF can only be
characterised from archived
stellar cluster images when a small ellipticity adjustment is applied.

The reason for the temporal variation of the ACS PSF is not fully understood.
Figure~\ref{fig:VzPSFvariation} shows that the F606W PSF
becomes more circular $(|\varepsilon| 
\rightarrow 0)$ in time, in contrast to the F850LP PSF which 
becomes more elliptical. This could potentially
be explained by a slow de-focus if one also considers
the poorly understood strong horizontal diffraction spike seen in the F850LP
PSF.  It is unlikely that this spike is caused by the ACS optics and it will
therefore remain unaffected by any de-focus. A de-focus will circularise the
images, as seen from the F606W data, and lower the contrast between the PSF
core and diffraction 
spike in the F850LP data, increasing the F850LP $\varepsilon^*_1$ component. 
 
\begin{figure} 
\centerline{\epsfig{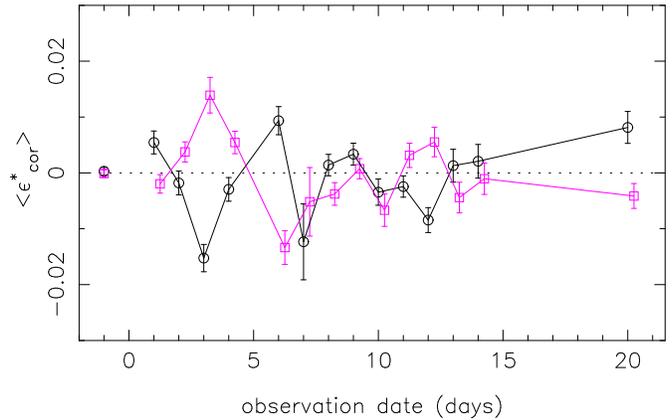}} 
\caption{The F606W corrected stellar ellipticity parameters $\varepsilon_1^{
  {\rm cor} *}$ (circles) and $\varepsilon_2^{ {\rm cor} *}$ (squares)
  averaged across the full ACS field of 
  view as a function of observation date.  The corrected stellar 
  ellipticity averages to zero over the whole data set as shown by the
  isolated points on the far left of the plot. 
  In general the PSF correction works well reducing the corrected ellipticity
  as a function of observation date close to zero, and significantly reducing
  the scatter (i.e removing the anisotropy) seen in
  Figure~\ref{fig:VzPSFvariation} (plotted on a different scale).  There is
  however  
  the occasional case where the stellar ellipticity is significantly different
  from zero showing that the semi-time-dependent modelling is only a partial
  solution for time varying PSF modelling.
  The errors plotted are the errors on the mean corrected stellar ellipticity.
} 
\label{fig:correctedPSFvariation} 
\end{figure}

Figure~\ref{fig:PSF} shows that our semi-time-dependent PSF
correction reduces the average stellar
ellipticity to zero and this generally holds when we measure the PSF
corrected stellar ellipticity as a function of observation date, as shown in
Figure~\ref{fig:correctedPSFvariation} . There are however notable
exceptions, for example ACS images taken on the third and sixth observation
days, showing that the semi-time-dependent modelling is only a partial
solution for time varying PSF modelling.

\begin{figure} 
\centerline{\epsfig{file=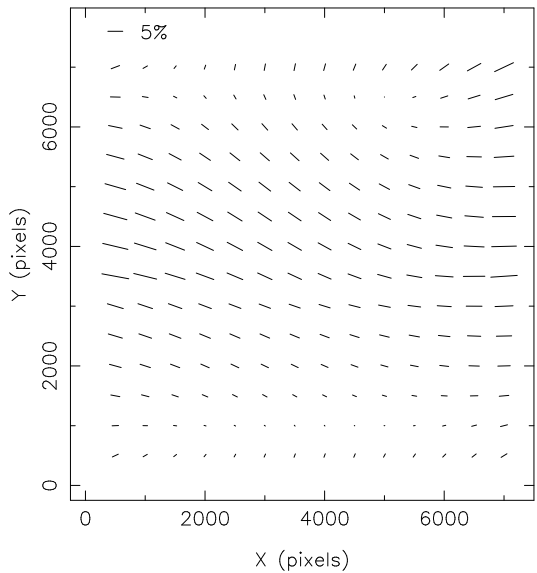,width=7.5cm,angle=270,clip=}} 
\caption{The difference between stellar ellipticity calculated from PSF
models for each of the two GEMS data sets with $r_g = 5.9$ pixels.
This shows that PSF variation is not a simple change in average ellipticity
but also an instability in the PSF anisotropy across the field of view,
varying at the maximum level of $\Delta \varepsilon^* = 5$\%.} 
\label{fig:PSF_time1_time2} 
\end{figure}

To investigate the effect on the anisotropy of the PSF
Figure~\ref{fig:PSF_time1_time2} compares the difference between the
stellar ellipticity predicted by the PSF models
calculated for the two GEMS data sets (days 1-10 and days 11-20). To convert
$\Delta p$ into a difference in 
stellar ellipticity we multiply by the average stellar smear polarisability
$P^{\rm sm*}$ 
following equation (\ref{eqn:ecor}).  This comparison reveals variation in the
anisotropy of the PSF at
the level of, at maximum, $\Delta \varepsilon^* \sim 5$\%.  We hence conclude
that the PSF time variation is not simply a 
change in average ellipticity but rather
an instability in the PSF anisotropy across the field of view.

Creating PSF models for each ACS tile based on at maximum a few tens of
stellar objects will yield systematic errors larger than the
variation in the mean corrected stellar ellipticity that we see in
Figure~\ref{fig:correctedPSFvariation}.  Thus, the short-term PSF variation
cannot be simply modelled with the method that
we have used.  The time 
dependent variation of the PSF whilst significant
is at a low level and we therefore proceed with
our semi-time dependent PSF model noting that short term variation in the
PSF may well contribute to systematic errors (see
Section~\ref{sec:shear_power}).

\subsection{Isotropic correction}

The application of the anisotropic PSF correction to observed galaxy
ellipticities, 
through equation (\ref{eqn:ecor}), leaves 
an effectively isotropic distortion.
This distortion makes objects rounder as a
result of both the PSF and the Gaussian weight function used to
measure the galaxy shapes, affecting smaller galaxies more strongly than the
larger galaxies.  To correct for this effect and convert weighted
galaxy ellipticities $\varepsilon$ into unbiased shear estimators
$\hat{\gamma}$, we use the pre-seeing shear polarisability tensor $P^\gamma$,
equation (\ref{eqn:Pgamma}).  $P^\gamma$ is calculated for each galaxy from the
measured galaxy smear and shear polarisability tensors, $P^{\rm sm}$
and $P^{\rm sh}$, and a term that is dependent
on stellar smear and shear polarisability tensors; $\left(P^{\rm sm *}
\right)_{\mu \delta}^{-1} P^{\rm sh *}_{\delta \beta}$.  As the $P^\gamma$
correction is isotropic we can calculate this stellar term purely 
as a function of smoothing scale $r_g$, 
averaging over all the stellar objects that were used in the previous
anisotropic PSF modelling.  Note that we calculate a different value for each
data set as the PSF variation we see may well be related to camera focus which
will effect the $P^\gamma$ correction 
as well as the anisotropic PSF correction.  $P^\gamma$ is a very noisy 
quantity, especially for small galaxies, but as we expect there to be no
difference in the $P^\gamma$ correction for the $\gamma_1$ and $\gamma_2$
shear components, we can reduce this noise somewhat by treating $P^\gamma$ as
a scalar equal to half its trace (note that the off diagonal terms of
$P^\gamma$ are typically very small).  

In an effort to reduce the noise on $P^\gamma$ still
further, $P^\gamma$ is often fit as a function of $r_g$
\citep{HFKS98,BMRE,MLB02,Massey}.  With space-based data this fitting method 
produces a bias as $P^\gamma$ is a strong function of galaxy ellipticity where
the dependence can be demonstrated by considering that in the case of 
a galaxy observed in the absence of PSF
smearing and shearing, $P^\gamma$ reduces to $P^\gamma =
2(1-e^2)$, where $e$ is the unweighted galaxy ellipticity.
$\gamma =\varepsilon/P^\gamma $ is very sensitive to small errors in a
functional fit 
of $P^\gamma(r_g,\varepsilon)$ and we therefore do not use any form of fitting
to $P^\gamma$.  Although this decreases the signal-to-noise of the shear
measurement, it avoids any form of shear calibration bias which would
not be identified with an E/B mode decomposition as discussed in 
Sections~\ref{subsec:EB} and~\ref{sec:shear_variance}. 

\subsection{Catalogue selection criteria}

For our final PSF corrected shear catalogue we select galaxies with size
$r_g>2.4$ pixels, galaxy shear $|\gamma|<1$, magnitude $21<{\rm m}_{606}<27$,
and ${\rm SNR} >15$. We remove galaxies from the catalogue with 
neighbouring objects closer than $20$ pixels ($0.6$ arcsec) to reduce noise
in the 
ellipticity measurement from overlapping isophotes.  These selection criteria 
yield 47373 galaxies in the F606W images, a number density of $\sim 60$
galaxies per square arcminute, and 23860 galaxies in the F850LP images and a
number density of $\sim 30$ galaxies per square arcminute.  Note $\sim 15\%$ of
our F606W sources have photometric redshift estimates from COMBO-17.  We find 
no significant correlations of the galaxy shear with chip position, galaxy
size, magnitude or SNR.   

To analyse the full GEMS mosaic we rotate the shear measurements from each ACS
tile into a common reference frame by
\be 
\left( 
\begin{array}{c} 
\gamma_1^{\rm rot} \\ 
\gamma_2^{\rm rot} 
\end{array} 
\right) = 
\left( 
\begin{array}{cc} 
\cos 2\phi & \sin 2\phi  \\ 
-\sin 2\phi & \cos 2\phi 
\end{array} 
\right) 
\left( 
\begin{array}{c} 
\gamma_1 \\ 
\gamma_2 
\end{array} 
\right), 
\label{eqn:rotg1g2}
\ee 
where $\phi$ is defined to be the 
angle between the $x$ axis of each ACS tile and a line of constant
declination.

\section{Comparison of F606W and F850LP data}
\label{sec:compVz}

\begin{figure} 
\centerline{\epsfig{file=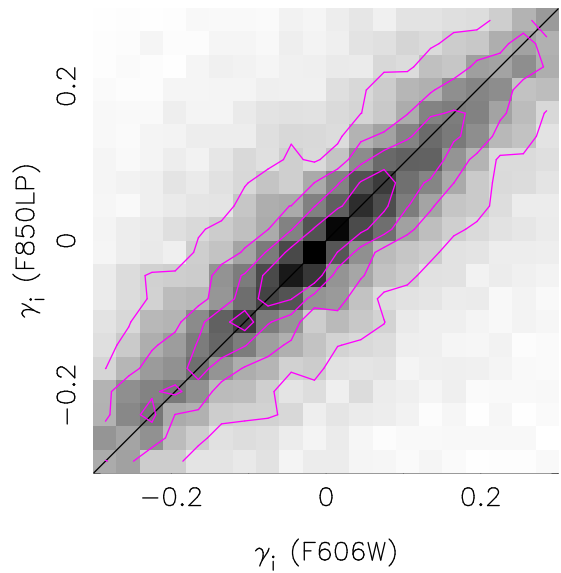,width=7.5cm,angle=270,clip=}} 
\caption{Comparison of galaxy shear $\gamma_i$ measured in both F606W and
   F850LP GEMS 
  imaging.  The grey-scale shows the number density of objects which
  cluster at low shear values (black $= 210$ galaxies, white $\rightarrow 0$
  galaxies).  Over-plotted are the grey-scale contours which 
  follow the
  1:1 relationship that we would wish to see between the two data sets.}
\label{fig:Vzcomp} 
\end{figure}

As the F850LP data is significantly shallower than the F606W data we
omit it from our cosmic shear analysis, but it is interesting to use the
galaxies detected in both F606W and F850LP as a consistency check to test if
our method is sensitive to the differences in filter PSFs seen in
Figure~\ref{fig:Vz_core_wings}.  Even though we expect galaxy morphology to
appear differently in the F606W and F850LP images, we expect our shear
estimates 
to remain consistent. Figure~\ref{fig:Vzcomp} shows the very good
agreement between galaxy shear measured in the F606W and F850LP images
where the grey-scale shows the
number density of objects.  This comparison shows that our method of measuring
shear produces very consistent results for galaxies imaged with different PSFs
and different noise properties, showing no significant calibration biases.  
In principle one should correct the galaxy ellipticity based on galaxy colour
because of the chromatic anisotropy of the PSF, but this comparison also shows
that the colour of the PSF does not significantly impact on the
shear measurement.

\section{GEMS redshift distribution}
\label{redshift_dist}
To interpret a weak lensing signal we need to know the redshift distribution
of the lensed sources (see equation (\ref{eqn:W})).  The deeper the survey is,
the stronger the signal we expect to measure.
We estimate the source redshift distribution of GEMS based on photometric
redshifts from the CDFS COMBO-17
survey \citep{C17} and spectroscopic redshifts 
from the CDFS VVDS survey \citep{VVDS}, by assuming that 
a magnitude dependent redshift distribution can be parameterised as
\be
n(z,{\rm mag}) \propto z^2 \exp \left[ - \left( \frac{z}{z_0({\rm mag})}
  \right)^{1.5} \right] 
\label{eqn:n_of_z}
\ee
where $z_0$ is calculated from the median redshift $z_m$ with $z_0 = z_m /
1.412$ \citep{BaughEfst}.
We calculate $z_m({\rm  mag})$ for each survey in magnitude 
bins of width 0.5 magnitude, out to the limiting magnitude of each survey
VVDS, ${\rm I}_{\rm AB} < 24$; COMBO-17, ${\rm R}_{\rm Vega} < 24$.  These
estimates are taken as lower limits for the true median redshift, as
both surveys suffer from redshift incompleteness at faint magnitudes.  
To calculate upper limits for the true median redshift we follow \citet{MLB02}
assuming those galaxies without an assigned redshift are most likely to be at
a higher redshift.  We place the percentage of galaxies without redshift
measurements at $z=\infty$ and re-calculate $z_m({\rm  mag})$, taking
our final median redshift estimate to be the midpoint between this upper limit
and the measured lower limit. In the cases
where the difference between our upper and lower limit constraints are larger
than the error on the mean of the distribution $\sigma_z/\sqrt{N}$ the  
uncertainty on the median redshift is given by the upper and lower limits.  For
bright magnitudes where the redshift measurements are fairly complete and the
number density of objects are relatively small, we place a
statistical uncertainty on the median redshift given by
$\sigma_z/\sqrt{N}$. For the COMBO-17 median redshift error 
we include the additional error on the photometric redshift estimate
where $\delta z / (1+z) \sim 0.02$ for ${\rm R}_{\rm Vega} < 22$ and
$\delta z / (1+z) \sim 0.05$ for $22<{\rm R}_{\rm Vega} < 24$.

\begin{figure} 
\centerline{\epsfig{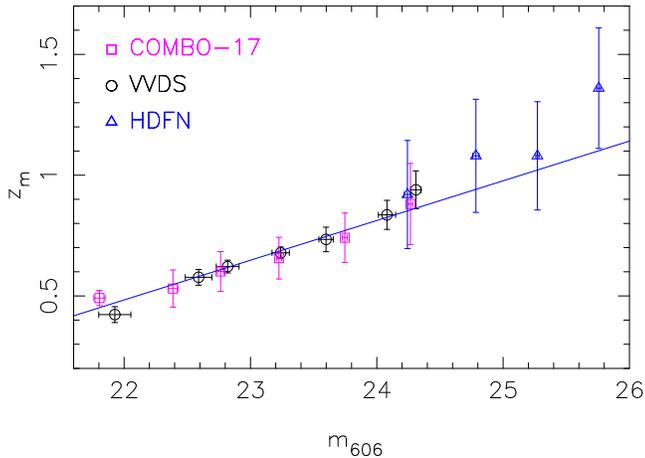}} 
\caption{The median galaxy redshift as a function of ${\rm m}_{606}$
 magnitude based on photometric redshifts from COMBO-17 (squares) and the HDFN
 (triangles), and spectroscopic redshifts from VVDS (circles).
Median redshift errors include uncertainty in redshift completeness which
dominates the COMBO-17 and VVDS results at faint magnitudes, and statistical
 uncertainty  
which dominates in the smaller galaxy samples at bright magnitudes.  
For the photometric redshifts we also include the average redshift error, added
 in quadrature. Magnitude 
errors show the uncertainty in the mean ${\rm m}_{606}$  magnitude
in each magnitude bin selected in ${\rm I}_{\rm AB}$ (VVDS), ${\rm
  R}_{\rm Vega}$ (COMBO-17) and ${\rm m}_{606}$ (HDFN).  Over-plotted is the
 best linear fit to the COMBO-17 and VVDS data.  The HDFN data is
 shown at faint magnitudes to justify our extrapolation to faint magnitudes 
of the COMBO-17/VVDS linear fit. } 
\label{fig:zm_VAB} 
\end{figure}

To convert the $z_m({\rm  mag})$ from the different surveys to $z_m({\rm
  m}_{606})$ we match the COMBO-17 and VVDS sources to the  
CDFS 5 epoch GOODS ${\rm m}_{606}$  catalogue \citep{GOODS}.  Note
that matching the comparatively shallow
redshift catalogues with deep 5 epoch GOODS data ensures that we are not 
subject to incompleteness in the ${\rm m}_{606}$ data.
We then calculate the mean
${\rm m}_{606}$ in each ${\rm I}_{\rm AB}$ and ${\rm R}_{\rm Vega}$
magnitude bin, assigning an uncertainty in the ${\rm m}_{606}$ 
magnitude of each bin given by $\sigma_{606}/\sqrt{N}$.
Figure~\ref{fig:zm_VAB} shows the 
combined results from both surveys which are in very good agreement, 
and the best linear fit;
\be
z_m = -3.132 + 0.164\, {\rm m}_{606} \,\,\,\,\,\,\, (21.8 <{\rm m}_{606}
< 24.4).
\label{eqn:zmvab}  
\ee
To estimate the median redshift of our galaxy sample fainter than ${\rm
  m}_{606} = 24.4$ we extrapolate the above relationship.  This is justified
by the $z_m:{\rm m}_{606}$ relationship measured in the
Hubble Deep Field North (HDFN) \citep{Lanzetta,Fsoto} shown (triangles) in 
Figure~\ref{fig:zm_VAB}, where the median redshift error combines the
photometric redshift error $\delta z  / (1+z) \sim 0.1$ \citep{Fsoto} and the
statistical uncertainty in each bin.
Note that only the ${\rm m}_{606}>24$ points are 
shown for clarity, but there is also good agreement between COMBO-17, VVDS and
HDFN at brighter magnitudes

We estimate the redshift distribution for GEMS $\phi(z)$ through
\be
\phi(z) = \sum_{i=1}^{M} N(i) n(z,{\rm m}_{606}(i)) / \sum_{i=1}^{M}
N(i) 
\ee
where we bin the GEMS sources into $i=1,M$ magnitude slices of mean
magnitude ${\rm m}_{606}(i)$, where each bin contains $N(i)$
galaxies.  $n(z,{\rm m}_{606}(i))$ is calculated through
equation (\ref{eqn:n_of_z}) with $z_0({\rm m}_{606}) = z_m({\rm m}_{606})/1.412$ as estimated from equation (\ref{eqn:zmvab}).  The 
calculated $\phi(z)$ is very similar to a magnitude independent 
$n(z)$ equation (\ref{eqn:n_of_z}), with $z_m =1.0$ 
and as such, for simplicity when
deriving the weak lensing theoretical models in the following analyses, we
will use a magnitude independent $n(z)$ with 
$z_m =1.0 \pm 0.1$, where the error given derives from the accuracy of
the $z_m({\rm m}_{606})$ fit, shown in Figure~\ref{fig:zm_VAB}.

\section{Analysis: 2-point statistics}
\label{sec:analysis_cs}
In this section we use GEMS F606W data to 
measure the shear correlation function, the shear 
variance statistic and the shear power spectrum 
performing several diagnostic tests for systematic errors.  We also determine
an additional sampling error in order to account for the fact that GEMS
images only one field.

\subsection{Jackknife Method} 
\label{sec:jackknife}
In the analysis that follows we will often make use of the jackknife
statistical 
method (see for example \citealt{Wallstats}) to obtain correlation functions
with a robust estimate of the covariance matrix.  The algorithm is quite
simple, if a little time 
consuming.  We are interested in the two-point correlation
function which we first calculate from
the whole survey and write as a data vector
$C=C(\theta_1,\theta_2,......)$.   
We then divide our sample into $N$ separate sub-regions
on the sky of approximately equal area and calculate the correlation
function $C_l= C_l(\theta_1,\theta_2,......)$, 
$l=1..N$ times omitting one sub-region in each calculation.  Note that for a
traditional jackknife we would perform the measurement $N=N_{\rm galaxy}$
times removing a 
single galaxy each time, but this is computationally prohibited and provided
$N$ is larger than the number of angular bins, this modified jackknife method
is valid \citep{SDSSjk}.
Defining 
\be
C_l^* = NC - (N-1)C_l\, ,
\ee
the jackknifed estimate of the
correlation function, $\hat{C}$,
is then given by the average $\hat{C} = \langle C_l^*\rangle$. 
The jackknife estimated statistical
covariance of the correlation function $C(\theta_i)$ in angular bin $i$
and the correlation 
function $C(\theta_j)$ in angular bin $j$ is given by 
\ba
\lefteqn{\lgl \Delta C(\theta_i) \Delta C(\theta_j) \rgl = \frac{1}{N(N-1)}
  \times} \nn 
&&\sum_{l=1}^{l=N}  
\left(C_l^{*}(\theta_i) - \hat{C}(\theta_i)\right)
\left(C_l^{*}(\theta_j) - \hat{C}(\theta_j)\right).
\label{eqn:jkcov}
\ea

\subsection{Tests for PSF contamination}
\label{sec:PSFtests}

In this section we perform several diagnostic tests to determine whether 
residual PSF-related correlations remain after the PSF correction of
Section~\ref{sec:PSF}. Measuring the magnitude  
of PSF contamination as a function of angular scale enables us to
determine which angular scale shear correlations are free from sources of
systematic errors and are therefore useful for cosmological parameter
estimation. 

\subsubsection{Star-galaxy cross correlation}
\label{subsec:csys}
\begin{figure} 
\centerline{\epsfig{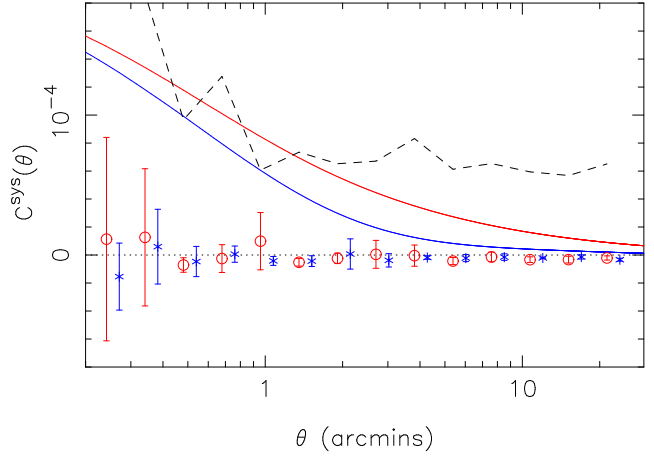}} 
\caption{Star-galaxy cross correlation functions 
$C_{tt}^{\rm sys}$ (circles) and $C_{rr}^{\rm sys}$ (crosses) compared to 
  theoretical galaxy-galaxy shear correlation functions $\langle
  \gamma_{r}\gamma_{r} \rangle_\theta$ (upper curve) and $\langle
  \gamma_{t}\gamma_{t} \rangle_\theta$ (lower curve) with $z_m = 1.0$,
  $\Omega_m = 0.3$, and $\sigma_8 = 0.7$.  For comparison we also measure 
  the star-galaxy
  cross correlation in the absence of PSF corrections, shown dashed, where for
  clarity we plot $\frac{1}{2}( 
  C_{tt}^{\rm sys} + C_{rr}^{\rm sys})$.}   
\label{fig:csys} 
\end{figure}

\citet{BMRE} show that residual PSF-related distortions
add a component $C_{ij}^{\rm sys}$ to the measured
correlation function $\langle\gamma_i\gamma_j\rangle$ where
\be
C_{ij}^{\rm sys} = \frac{\langle
\gamma_i \varepsilon_j^*\rangle \langle \gamma_j 
\varepsilon_i^*\rangle}{\langle \varepsilon_i^*
\varepsilon_j^*\rangle}. 
\label{eqn:Csys}
\ee
This method to test for residual PSF contamination in the data is similar to
the cell 
averaged test described in Section~\ref{sec:PSF} but in this case we
look for 
correlations as a function of angular scale, not chip position, thereby
revealing any PSF time variation effects that remain after our
semi-time-dependent PSF correction has been applied. 
We estimate $C_{ij}^{\rm sys}$ and associated errors 
using the modified jackknife method detailed
in Section~\ref{sec:jackknife}, with $\varepsilon_i^*$ determined at
$r_g = 5.9$, the median galaxy size in the survey.  We use $N=25$ sub-regions
in the jackknife estimate of $M=14$ logarithmic angular bins from $\theta =
0.2$ arcmin to $\theta=25$ arcmin.  Figure~\ref{fig:csys}
shows the resulting star-galaxy cross correlation functions 
$C_{tt}^{\rm sys}$ and $C_{rr}^{\rm sys}$, compared to theoretical
galaxy-galaxy shear correlation functions with $\Omega_m = 0.3$, and $\sigma_8
= 0.75$.  We find that the star-galaxy cross
correlation is consistent with zero indicating that the measurement of 
galaxy-galaxy shear correlations from the GEMS data 
will be free from major sources of systematics.  For comparison we also
measure the star-galaxy cross correlation when we have not included a
correction for the distorting PSF.  This reveals 
a correlation signal (shown dashed) that exceeds the weak cosmological signal
that we 
wish to measure, stressing the importance for a good understanding of the
PSF.

\subsubsection{E/B mode decomposition of shear correlations}
\label{subsec:EB}
An alternative diagnostic to determine the level of systematic 
errors is to decompose 
the shear correlation function into `E-modes' and `B-modes'
\citep{CNPT02}.
Weak gravitational lensing produces gradient curl-free distortions (E-mode), 
and contributes only to the curl distortions (B-mode) at small angular 
scales, $\theta < 1$ arcmin, 
due to source redshift clustering \citep{SchvWM02}. 
A significant detection of a B-type signal in weak lensing surveys is 
therefore an indication that ellipticity correlations exist  
from residual systematics within the data and/or from intrinsic galaxy 
alignments (see \citet{HBH04} for a discussion of the latter). 

Defining the sum and difference
of the tangential and rotated correlation functions,
\be 
\xi_{\pm}(\theta) = \langle \gamma_t \gamma_t \rangle_{\theta} 
\pm \langle \gamma_r \gamma_r \rangle_{\theta} ,
\label{eqn:xiplusminus}
\ee
\citet{CNPT02} show that the shear correlation functions can be decomposed
into the following E- and B-type correlators, 
\be
\xi^E(\theta)=\frac{\xi_\myplus(\theta)+\xi'(\theta)}{2}\ \ \ \ \ \ 
\xi^B(\theta)=\frac{\xi_\myplus(\theta)-\xi'(\theta)}{2}
\label{eqn:xieb}
\ee
where
\be
\xi'(\theta)=\xi_\myminus(\theta)+4\int_\theta^\infty \frac{d\vartheta}{\vartheta} \xi_\myminus(\vartheta)
        -12\theta^2 \int_\theta^\infty \frac{d\vartheta}{\vartheta^3}\xi_\myminus(\vartheta).
\label{eqn:xipr}
\ee
Our data extends at maximum to $\theta = 28$ arcmin necessitating the use of
a fiducial cosmological model to complete the integral.
This prevents cosmological parameter estimation directly
from the E-modes, but as the variation in the sum of the 
model dependent part of the
integral is small $\sim 10^{-5}$, 
this method is still a valid diagnostic test for residual systematics within
the data.   Note that the mass aperture statistic E/B
decomposition \citep{Sch98} does not suffer from this limitation as the
integral over the shear correlation function spans from $0 \rightarrow
2\theta$.  This integral range does however introduce its own problems (see
\citealt{Massey} for a discussion) and limits the analysis to small scale power
\citep{vWb04}.  We therefore choose to use the E and B type 
correlators purely to test for B-type systematic errors within our data,
although see \citet{vWb04} for a method that uses the mass aperture statistic
to calibrate $\xi^E(\theta)$ for cosmological parameter estimation.
\begin{figure}
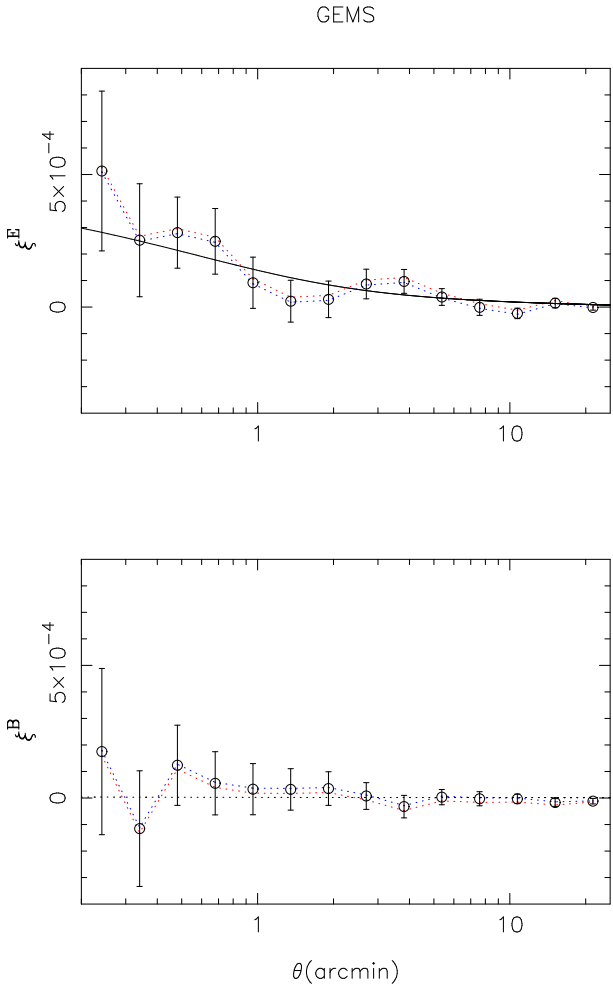
 
\centerline{\epsfig{file=GEMS_E_Crit.ps,width=6.5cm,angle=270,clip=}}
\centerline{\epsfig{file=GEMS_B_Crit.ps,width=6.5cm,angle=270,clip=}} 
\caption{E/B mode decomposition of the shear correlation function measured
  from the GEMS observations. 
The upper panel shows $\xi^{\rm bin}_E$ and the fiducial $\Lambda{\rm CDM}$
  theoretical  $\xi^{E}(\theta)$ model where $\sigma_8 = 0.7$ and
  the 
  median source redshift $z_m = 1.0$.  The lower panel shows $\xi^{\rm bin}_B$
  that is consistent with zero on all angular scales and can be compared to the
  theoretical model $\xi^{E}(\theta)$ shown dotted.  Using 
a different fiducial $\Lambda{\rm CDM}$ cosmological model to calculate
  $\xi^{\rm bin}$ has a small effect
(at the level of $10^{-5}$) which can be seen from the dotted curves where
  we have assumed  
$\sigma_8 = 0.6$ (lower curve E-mode, upper curve B-mode) and $\sigma_8 = 1.0$
  (upper curve E-mode, lower curve B-mode)} 
\label{fig:GM_EBmode} 
\end{figure}


Following \citet{PenWM} we define the $2n$ element 
vector $\xi_{i}= (\xi_\myplus(\theta),\xi_\myminus(\theta))$ which
we compute from GEMS, binning the data finely into $n=2000$ intervals
of $0.9$ arcsec (equivalent to a separation of $\sim 30$ ACS pixels).
The E/B correlators are then given by a $2n$ element vector
\be
\xi^{EB}_i=(\xi^E(\theta_i),\xi^B(\theta_i)) = T_{ij}\xi_j
\label{eqn:transform}
\ee
where $T$ is a $2n^2$ transformation matrix defined by
equation (\ref{eqn:xipr}).  To reduce noise 
we re-bin $\xi^{EB}_i$ into $M=14$ logarithmic
intervals from $\theta = 0.2$ arcmin to $\theta = 25.0$, represented by an $M
\times 2n$ projection operator $\xi^{\rm bin}_k=P_{ki}\xi^{EB}_i$.  

To calculate errors on $\xi^{\rm bin}_k$ we first calculate the covariance
matrix of 
the raw correlators $L_{ij} = \lgl \Delta \xi_i \Delta \xi_j \rgl$ using the
jackknife method detailed in Section~\ref{sec:jackknife}.  We find, in contrast
to \citet{PenWM}, that our slightly wider correlation bins ($\Delta \theta =
0.9$ arcsec compared to $\Delta \theta =0.6$ arcsec) are
correlated.  Note that the modified jackknife method in principle requires
$N>n$ sub-regions to calculate $L_{ij}$, but as we re-bin
$L_{ij}$ into $M=14$ broader angular scales to estimate the final errors on
$\xi^{\rm bin}$ our jackknife method, which is computationally 
time limited to $N=100$, is
still valid. 
The binned covariance matrix of the E/B
correlators is given by \citep{PenWM},
\be
L^{\rm bin}_{lm}= P_{li} T_{ij} L_{jk} 
T_{ko} P_{om},
\label{eqn:noiseb}
\ee
where the errors on $\xi^{\rm bin}_k$ are given by $\sqrt{L^{\rm bin}_{kk}}$.

We investigate E- and B-mode correlations in the GEMS 
and GOODS data separately with the measurement from the GEMS data shown in
Figure~\ref{fig:GM_EBmode}.  For this E/B mode decomposition we
have used a fiducial $\Lambda{\rm CDM}$ cosmological model to complete the
integral of equation (\ref{eqn:xipr}) with $\sigma_8 =
0.7$ and with a median source redshift for our galaxies of $z_m = 1.0$.  Using
a different fiducial $\Lambda{\rm CDM}$ cosmological model has a small effect
(at the level of $10^{-5}$) which can be seen from the dotted curves in
Figure~\ref{fig:GM_EBmode} where we have assumed 
$\sigma_8 = 0.6$ and $\sigma_8 = 1.0$.
We find that the E-modes are in good agreement with the fiducial cosmological
model and that the B-modes for GEMS are consistent with zero on all scales
$\theta>0.2$ arcmin. 
Whilst finding this result very encouraging we note that our B-modes
measured at small angular scales are very strongly correlated and 
noisy such that in the worse-case 
scenario, represented by the upper end of the error bars shown, the B-modes
exceed the signal that we wish to measure.  This motivates our desicion to
conservatively limit our shear correlation function analysis to angular scales
$\theta>0.65$ arcmins within the GEMS data where we can be confident that the
signal we measure is cosmological and not systematic.
 
The PSF model for the GOODS data is determined from approximately half 
the number of stars which were used to determine the two
semi-time-dependent GEMS PSF models, as 
GOODS spans approximately one quarter of the area of the GEMS observations.  
We would therefore expect to find a poorer PSF correction with the GOODS data
which is seen with the presence of non-zero B-modes
at angular scales $\theta<1$ arcmin.  We therefore only include GOODS data
in our shear correlation function analysis for angular scales $\theta>1$
arcmin. 

\subsection{The shear correlation function}
\label{sec:ctheta}

Having shown in the previous section
that we are not contaminated by significant non-lensing correlations, 
we can now measure the 
weak lensing shear correlation function $\langle
\gamma(\bm{\theta})\gamma(\bm{\theta} + \Delta \theta) \rangle$.
To obtain results that are independent of the initial frame of reference we
measure the
tangential and rotated shear correlation functions, equation (\ref{eqn:gtgt})
and equation (\ref{eqn:grgr}) respectively. These can be estimated from the
data by 
\be
E[\gamma_{ _r^t} \gamma_{ _r^t}]_\theta = \frac{1}{\rm Npairs} \sum_{\rm pairs}
\gamma_{ _r^t} (\bx) \, \gamma_{ _r^t} (\bx + \bm{\theta}),
\label{eqn:tanrad}
\ee
where the tangential shear $\gamma_t$ and rotated shear $\gamma_r$ are 
given by equation (\ref{eqn:rotg1g2}). 
$\gamma_t=\gamma_1^{\rm rot}$ and $\gamma_r = \gamma_2^{\rm rot}$ where the
rotation angle $\phi$ is now 
defined to be the angle between the $x$ axis and the line joining each galaxy
pair. Note this rotation follows the initial rotation that sets the full
GEMS shear catalogue into the same reference frame.
We also calculate the cross-correlation function $E[\gamma_{t}
  \gamma_{r}]_\theta$, which the parity invariance of weak lensing predicts
to be zero.

\begin{figure} 
\centerline{\epsfig{file=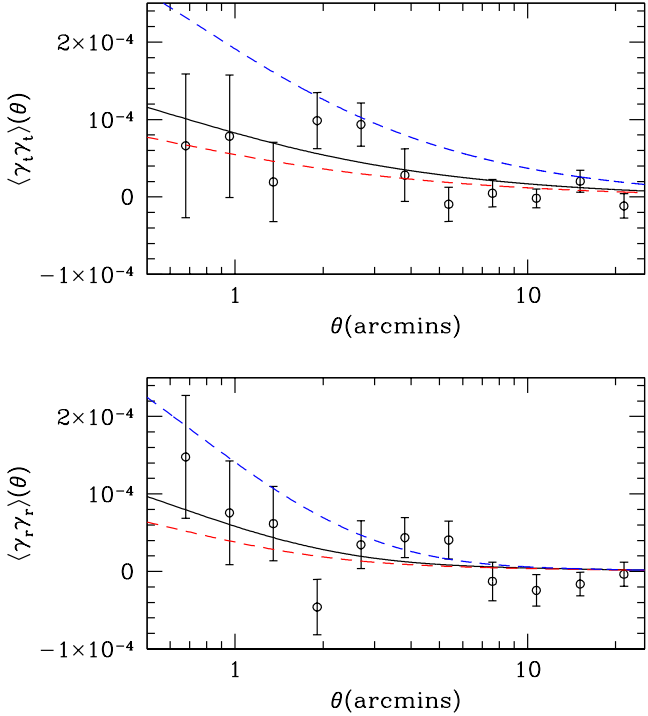,width=8.5cm,angle=0,clip=}} 
\caption{Shear correlation functions  $E[\gamma_{t} \gamma_{t}]_\theta$ (upper
  panel) and $E[\gamma_{r} \gamma_{r}]_\theta$ (lower panel) estimated from the
  GEMS data using a 
  modified jackknife technique.  The GOODS section of the GEMS survey is only
  included in measurements of $E[\gamma\gamma]_\theta$ for $\theta > 1.0$
  arcmin so as not to include the non-lensing B-mode systematics seen in the
  GOODS data at smaller angular scales. Over-plotted are theoretical $\Lambda
  CDM$ models $\langle \gamma_{t} \gamma_{t} \rangle_\theta$ (upper panel) and
  $\langle \gamma_{r} \gamma_{r} \rangle_\theta$ (lower panel) with $\sigma_8
  = 0.6$ (dashed  
  lower), $\sigma_8 = 0.7$ (solid) and $\sigma_8 = 1.0$ (dashed upper).}
\label{fig:Ctheta} 
\end{figure}

We calculate the shear correlations using the modified jackknife method
detailed in Section~\ref{sec:jackknife}.  We use $N=25$ sub-regions
in the jackknife estimate of $M=11$ logarithmic angular bins from $\theta =
0.65$ arcmin to $\theta=25$ arcmin and we include data from the GOODS area
only for angular scales $\theta>1$ arcmin. 
This ensures that the shear correlation measurement at small angular scales
$\theta<1$ arcmin is 
not contaminated by the small-scale non-lensing distortions found in the GOODS
data.  
We show the resulting jackknife estimates for the tangential and rotated
shear correlation functions in
Figure~\ref{fig:Ctheta}. The theoretical $\Lambda CDM$ models over-plotted
are calculated from equation (\ref{eqn:gtgt}) and equation (\ref{eqn:grgr})
with a median 
galaxy redshift $z_m = 1.0$ and $\sigma_8 = (0.6,0.7,1.0)$.  Note that we find
the cross-correlation 
$E[\gamma_{t} \gamma_{r}]_\theta$ to be consistent with zero on all scales, as
expected, supporting our findings that we are not contaminated by significant
non-lensing distortions.

\subsubsection{Sampling variance}
\label{sec:sampvar}
The GEMS mosaic samples only one area of sky and as such our results are
subject to additional sampling variance errors.  To address this issue we have
created 100, $28 \times 28$ arcmin, 
Gaussian realisations of a shear power spectrum calculated for a
$\sigma_8 = 0.75$, $\Lambda$CDM cosmology for  
sources with $z_m=1$.  We measure the shear correlation function from each 
realisation, populating the shear field with 60 circular 
galaxies per square arcmin.  The variance we measure between the results from
each realisation then provides us with an estimate of the sampling
variance.  We calculate the sampling covariance matrix from the 100 independent
realisations and add this to the jackknifed covariance matrix measured from
the GEMS data, slightly overestimating the sampling variance error on
small scales. The reader should note that this method is somewhat 
cosmology dependent
but it is sufficient to estimate the amplitude of sampling
variance for our cosmological parameter constraints in
section~\ref{sec:cosmoparam}.   Future work will investigate the impact of 
non-Gaussianity on our sampling variance estimation using the cosmological 
ray-tracing N-body simulations of \cite{Whitesims}. 

\subsection{The shear variance statistic}
\label{sec:shear_variance}

Space-based lensing surveys to date and early ground-based lensing results
focused on the shear variance statistic, equation
(\ref{eqn:shearvar}), to analyse the data.  This statistic  
produces the
highest signal-to-noise measurement of weak lensing shear and can be estimated
from the data by splitting the sample into N circular 
cells of radius $\theta$ and calculating the shear variance in excess of noise
(see \citealt{MLB02} for a minimum variance estimator).  As discussed in
Section~\ref{subsec:EB} measuring the B-mode of the shear correlation function 
allows one to select angular scales above which one can be confident
that the presence of non-lensing distortions are insignificant.  Very small
scale systematic distortions are poorly understood and are sucessfully 
ignored by the shear correlation statistic.
For the shear variance statistic however, small
scale non-lensing distortions are included in the measurement at all angular
scales biasing the shear variance statistic.

To assess the impact of our residual small scale non-lensing distortions on
the shear variance statistic we can, in a similar fashion to
Section~\ref{subsec:EB}, 
decompose the signal into its E-mode and B-mode components.  \citet{SchvWM02}
show that the shear variance of the E- and B-mode can be obtained in terms of
the shear correlation functions $(\xi_\myplus, \xi_\myminus)$,
equation (\ref{eqn:xiplusminus}), through
\be
\langle \gamma^2 \rangle_\theta^E  = \int_0^\infty
\frac{d\vartheta\,\vartheta}{2\theta^2} \left[ \xi_\myplus(\vartheta)
  S_\myplus\left(\frac{\vartheta}{\theta}\right) + \xi_\myminus(\vartheta)
  S_\myminus\left(\frac{\vartheta}{\theta}\right) \right],
\label{eqn:top_hat_E}
\ee
\be
\langle \gamma^2 \rangle_\theta^B  = \int_0^\infty
\frac{d\vartheta\,\vartheta}{2\theta^2} \left[ \xi_\myplus(\vartheta)
  S_\myplus\left(\frac{\vartheta}{\theta}\right) - \xi_\myminus(\vartheta)
  S_\myminus\left(\frac{\vartheta}{\theta}\right) \right],
\label{eqn:top_hat_B}
\ee
where $S_\myplus$ and  $S_\myminus$ are given in equation (39) and equation
(42) of 
\citet{SchvWM02}.  Note that $S_\myminus$ does not cut off at finite
separation and as such one needs to include a fiducial cosmological model to
complete the integral, as in Section~\ref{subsec:EB}.  We calculate $\langle
\gamma^2 \rangle_\theta^E$ and $\langle \gamma^2 \rangle_\theta^B$ following  
the method of \cite{PenWM}, detailed in Section~\ref{subsec:EB}, where the
transformation matrix $T$ of 
equation (\ref{eqn:transform}) is now defined by equation (\ref{eqn:top_hat_E})
and equation (\ref{eqn:top_hat_B}).  
\begin{figure}
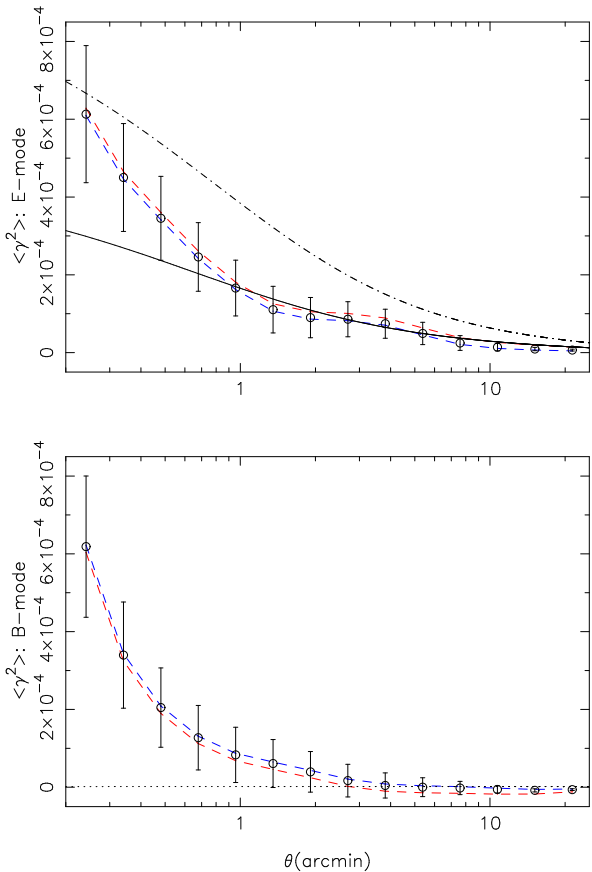
 
\centerline{\epsfig{file=top_hat_E.ps,width=5.75cm,angle=270,clip=}} 
\centerline{\epsfig{file=top_hat_B.ps,width=5.75cm,angle=270,clip=}} 
\caption{E/B mode decomposition of the shear variance.
The upper panel shows $\langle \gamma^2 \rangle_\theta^E$ and the
fiducial $\Lambda{\rm CDM}$ 
  theoretical $\langle \gamma^2 \rangle$ model (solid)
  where $\sigma_8 = 0.7$ and the 
  median source redshift $z_m = 1.0$.  The lower panel shows $\langle \gamma^2
  \rangle_\theta^B$ 
  that is consistent with zero on angular scales $\theta>1.5$ arcmin. Using 
a different fiducial $\Lambda{\rm CDM}$ cosmological model to calculate 
$\langle \gamma^2 \rangle_\theta^{EB}$ has a small effect 
(at the level of $3\times 10^{-5}$) which can be seen from the dotted curves
where we have assumed  
$\sigma_8 = 0.6$ (lower curve E-mode, upper curve B-mode) and $\sigma_8 = 1.0$
  (upper curve E-mode, lower curve B-mode).  The significant non-lensing B-modes
  at $\theta<1.5$ arcmin indicate residual small scale systematics that, with
  the shear variance statistic, are included at all angular scales. These
  non-lensing distortions also contribute to the measured E-mode making
the signal appear to favour a high value for $\sigma_8$
(upper panel: $\Lambda{\rm CDM}$  $\sigma_8 = 1.0$ theoretical model
over-plotted dot-dashed).}
\label{fig:top_hat_EB} 
\end{figure}

Figure~\ref{fig:top_hat_EB} shows the result of our E/B mode decomposition of
the shear variance statistic revealing significant B-modes on scales
$\theta<1.5$ arcmin. Replacing the small
scale $\xi_\myplus$ and $\xi_\myminus$ measurements 
($\theta<0.2$ arcmin) with a theoretical model, 
we find B-modes consistent with zero on all scales.  
The B-modes shown in Figure~\ref{fig:top_hat_EB} 
therefore result from very small
scale strong non-lensing distortions that bias 
the shear variance statistic even at larger angular
scales. 
Non-lensing distortions are likely to contribute equally 
to the measured E-mode making
the signal appear to favour a higher value for $\sigma_8$, when compared to
the large scale B-mode free shear variance measurements.  These large 
angular scales have previously been
unmeasureable from space-based data.  This demonstrates that 
in order to obtain reliable cosmological parameter constraints
it is vital to perform an E/B mode decomposition of the statistic 
that will be used to constrain cosmology.  We will not
use the shear variance statistic to constrain cosmological parameters, even on
B-mode free scales, preferring the shear correlation function statistic in
Section~\ref{sec:ctheta} and the shear power spectrum that we determine in the
following section.

\subsection{Shear power spectrum}
\label{sec:shear_power}

In addition to the shear correlation function and shear variance statistic of
the previous 
sections, we also quantify the two-point statistics of the shear
field by directly measuring its power spectrum, $P_\kappa$,
equation (\ref{eqn:Pkappa}). Power spectrum estimation from cosmological
data sets is a well-studied problem in the context of measuring the
statistical properties of the  
CMB (see \citealt{Efstathiou04} for an overview) and the methods developed in
this field are completely applicable to measuring power spectra from
weak lensing data sets. Here, we use a maximum likelihood estimator
(see for example \citealt{BJK}) to reconstruct the power spectrum of the shear 
field observed in the GEMS data. Our approach is based on the
prescription of \citet{HuWhite01} who proposed reconstructing the
three power spectra, $P^{\kappa\kappa}$, $P^{\beta\beta}$ and
$P^{\kappa\beta}$ as a series of step-wise `band-powers' where the
quantity $\ell(\ell+1)P^{ij}/2\pi$ is approximated as a constant within each
band. $P^{\beta\beta}$ is the power spectrum of the
B-modes while $P^{\kappa\beta}$ is the cross power-spectrum between
the E- and B-modes.   The maximum 
likelihood method automatically accounts for irregular survey
geometries, pixelization effects and produces error estimates, 
via the Fisher Information
matrix (see for example \citealt{TTH97}), which include sampling
variance and shot noise. \citet{HuWhite01} have tested the maximum
likelihood estimator on both Gaussian and N-body simulations, while
\citet{MLB02} have tested it on Gaussian simulations on scales similar to the 
GEMS data and have applied the estimator to the COMBO-17 weak lensing data set.
 
The maximum-likelihood decomposition
only necessitates the use of a fiducial
cosmological model to estimate the significance of the result.  This is in
comparison to 
the E/B correlators of Section~\ref{subsec:EB} and the E/B shear variance
measurements of Section~\ref{sec:shear_variance} where a fiducial
cosmological model is needed to complete integrals over the infinite
correlation function thereby invalidating their sole use for cosmological
parameter estimation.

To apply the estimator to the GEMS data, we bin the galaxy shear distribution
into $30\times30$ equal-size pixels of $\sim 1$ square arcmin. Writing this
pixelised shear distribution as a vector, ${\bf d}$, we then maximise
the likelihood function,
\begin{equation}
-2 \ln L({\bf C}|{\bf d}) = {\bf d}^t {\bf C}^{-1} {\bf d} + {\rm Tr}
 \, [ \ln {\bf C}],
\end{equation} 
using a Newton-Raphson scheme, as a function of the band-powers of the
three power spectra, $P^{\kappa\kappa}$, $P^{\beta\beta}$ and
$P^{\kappa\beta}$. Here, ${\bf C}$ is the data covariance matrix
which is the sum of the cosmological signal (equation (21) of \citet{MLB02})
and a noise term,  
\begin{equation}
{\bf N} = \frac{\gamma_{\rm rms}}{N_{\rm pix}} \,\, {\bf I},
\end{equation} 
where $\gamma_{\rm rms}$ and $N_{\rm pix}$ are the root mean square shear
and occupation number within each pixel respectively. The errors and
covariance of our final band-powers are approximated as the inverse
Fisher matrix, which is an excellent approximation provided that the
likelihood function is sufficiently Gaussian in the band-powers.    
 
\begin{figure} 
\centerline{\epsfig{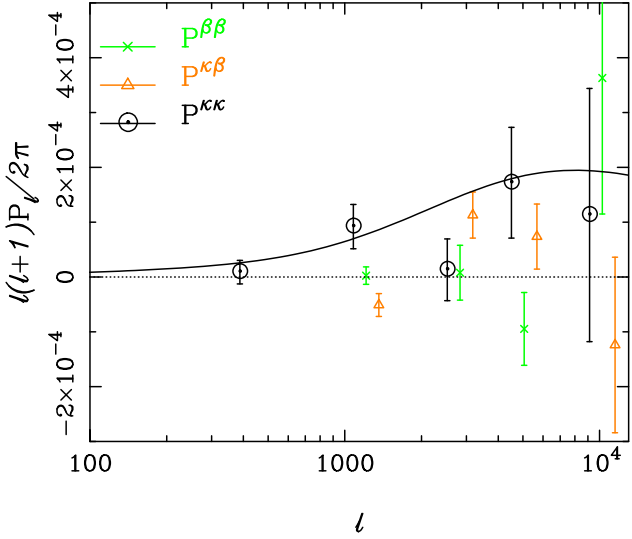}} 
\caption{The cosmic shear power spectra from GEMS.  Plotted on a linear-log
  scale are  $P^{\kappa\kappa}$ (circles), $P^{\beta\beta}$ (crosses) and
$P^{\kappa\beta}$ (triangles) in five band-averaged band-powers as a function
  of multipole, $\ell$.  The errors bars are estimated from the Fisher matrix and
  the $P^{\beta\beta}$ and $P^{\kappa\beta}$ have been slightly horizontally
  displaced for clarity.  The solid curve is the shear power spectrum
  estimated for a $\sigma_8 = 0.8$ normalised $\Lambda{\rm CDM}$ model.}
\label{fig:shearpow} 
\end{figure}

Figure~\ref{fig:shearpow} shows the results of applying the maximum likelihood
estimator to the GEMS data along with a theoretical shear power
spectrum for a $\Lambda$CDM model with $\Omega_m=0.3$ and
$\sigma_8=0.8$ with which we find reasonable agreement. The
measurements of the B-mode spectrum are mostly consistent with zero
although there is a significant detection of E-B cross-correlations on
medium scales.  We suspect that these come from the time-variation of the PSF
that we have only partially accounted for with our semi-time dependent PSF
modelling. 
Our PSF models are designed to decrease the average stellar ellipticity to
zero and therefore when averaging over the whole survey, as in the measurement
of the shear correlation function, the residual PSF
contamination is zero, as can be seen in Figure~\ref{fig:csys}. For the
shear 
power spectrum measurement however, where the field is decomposed into its
Fourier components, the time-dependent PSF contamination can be identified. 
Note that from the covariance of the $P^{\kappa\kappa}$ measurements 
we find that our band-power
measurements are almost independent of one another, apart from the slight
anti-correlation of neighbouring bands which is 
a natural consequence of the maximum likelihood estimator.


\section{Cosmological parameter estimation}
\label{sec:cosmoparam}

Having measured the 2-point statistics of the shear field within GEMS,
we can now compare these measurements with theoretical predictions in
order to place joint constraints on the matter density of the
Universe $\Omega_m$ and the normalisation of the matter power
spectrum $\sigma_8$. We do this using both our correlation function
measurements from Section~\ref{sec:ctheta} and the power spectrum estimates
from Section~\ref{sec:shear_power}. We use equation (\ref{eqn:Pkappa}) to
calculate our theoretical shear power  
spectra and equation (\ref{eqn:gtgt}) and equation (\ref{eqn:grgr}) to
calculate our 
theoretical correlation functions for a variety of cosmological
models. For these calculations we have used the transfer function of
\citet{HuEisen} for the dark matter power spectrum with an
initial power spectrum slope of $n=1$. To produce the non-linear power
spectrum from this, we use the fitting formulae of \citet{Rob}
and we fix $\Omega_m + \Omega_\Lambda = 1$. We also use the form of
equation (\ref{eqn:n_of_z}) for the input redshift distribution of source
galaxies. We consider 
models in the following regions of parameter space: $0.3 \le \sigma_8
\le 1.5$ ; $0.1 \le \Omega_m \le 1.0$ ; $64 \le H_0 \le 80 \,
{\rm km s^{-1} Mpc^{-1}}$ and $0.9 \le z_m \le 1.1$.
  
\begin{figure} 
\centerline{\epsfig{file=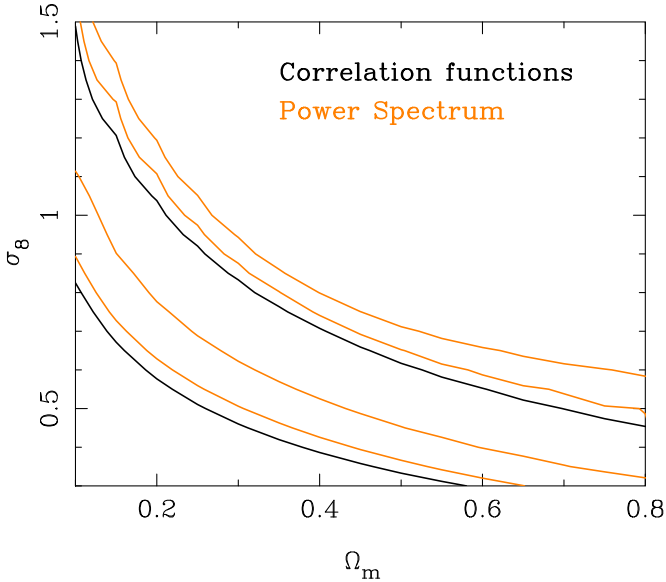,width=8.75cm,angle=0,clip=}} 
\caption{The likelihood surface of $\sigma_8$ and $\Omega_m$ from GEMS as
calculated using the shear correlation function (dark contour) and using
the shear power spectrum (light contours).  We plot the $1\sigma$ and
$2\sigma$ confidence regions for the shear power spectrum, and, for clarity, 
the $1\sigma$ confidence regions for the shear correlation function}
\label{fig:s8om} 
\end{figure}

Writing our correlation function measurements as a data vector, 
\begin{equation}
{\bf d} = \{ C_1(\theta_1),...,C_1(\theta_n), C_2(\theta_1),...,C_2(\theta_n) \},
\end{equation}
for each theoretical model, we calculate 
\begin{equation}
\chi^2=\left[ {\bf d} -{\bf x}\right]^T
{\bf V}^{-1} \left[ {\bf d} -{\bf x}\right],
\end{equation}
where ${\bf x} = {\bf x}(\sigma_8,\Omega_m,H_0,z_m)$ are the theoretical
correlation functions ordered in a similar manner to our data
vector. ${\bf V} = \langle{\bf dd}^T\rangle$ is the sum of the covariance
matrix of our correlation function measurements as
estimated from the data using equation (\ref{eqn:jkcov}) and a sampling
covariance matrix as detailed in section~\ref{sec:sampvar}.
The fitting of the power spectrum
measurements is done in a similar fashion where ${\bf V}$, the covariance
matrix of the band-power measurements, comes from a Fisher error analysis that
automatically includes sampling variance.
After calculating $\chi^2$
values for each of our theoretical models, we marginalise over the
Hubble constant, $H_0$ with a prior set by the $1^{\rm st}$ year WMAP results 
($H_0=72 \pm 5$ km s$^{-1}$ Mpc$^{-1}$; \citet{WMAP}). We also 
marginalise over the median redshift of the source galaxies, $z_m$
using $z_m=1.0 \pm 0.1$ as estimated in Section~\ref{redshift_dist}.

The resulting constraints in the $\sigma_8 - \Omega_m$ plane for both
the shear correlation function and shear power spectrum estimators are shown in
Figure~\ref{fig:s8om}. We find good agreement between the constraints obtained
using the two different measures: for the correlation function
measurements, we find
\begin{equation}
\sigma_8 (\Omega_m/0.3)^{0.65}=0.68 \pm 0.13
\end{equation}
while using the power spectrum analysis, we find a slightly higher
value of 
\begin{equation}
\sigma_8 (\Omega_m/0.3)^{0.65}=0.72 \pm 0.11.
\end{equation}

\section{Conclusion}
\label{sec:conc}

In this paper we have presented the detection of weak gravitational lensing by
large-scale structure in the GEMS survey, demonstrating that our shear
correlation signal is
uncontaminated by significant non-lensing shear distortions.  GEMS, imaged by
the ACS on 
HST, spanning $795$ square arcmin, is the largest contiguous space-based
mosaic that has undergone a cosmic shear analysis to date.  This has enabled
us to measure
cosmic shear over a large dynamic range of angular scales; from the small
scales ($\theta = 0.65$ arcmin) that are difficult to probe with
ground-based surveys, 
up to the large scales ($\theta = 21.0$ arcmin) that were previously
inaccessible to space-based surveys. 
Our careful analysis, where we have considered forms of selection bias,
centroid bias and calibration bias, geometric shear distortions and PSF
contamination, has yielded an 
unbiased\footnote{Our measurement is unbiased if we assume that the KSB+
  method applied provides us 
  with an unbiased estimate of galaxy shear $\gamma$, which has been shown to
  be true with ground-based data \citep{erben}.  The equal galaxy shear (on
  average)  
  measured in our F606W and F850LP data suggests that the impact of
  strongly non-Gaussian space-based PSFs on the KSB+ method is small,
  supporting its use as an unbiased shear estimator in this paper.  This
  will be investigated further with 
  sheared spaced-based image simulations in a forthcoming paper.} 
measurement of the shear correlation function
uncontaminated by non-lensing `B-mode' distortions.  This has allowed us to
set joint constraints uncontaminated by major sources of systematic errors
on the matter density of the
Universe $\Omega_m$ and the normalisation of the matter power
spectrum $\sigma_8$ finding $\sigma_8 = 0.73 \pm 0.13$ for WMAP constrained
$\Omega_m = 0.27$ \citep{WMAP}.
It is interesting to note that the GEMS cosmological parameter constraints are
very similar to 
those from the COMBO-17 survey \citep{MLB02,HBH04}, a deep multi-colour survey
which spans $\sim 4$ times the 
area of GEMS.  This results from the higher number density
of resolved galaxies in space-based data
and the higher signal-to-noise measurements of
galaxy shear which are achievable with higher resolution data (Brown et
al. in prep).

We have presented a thorough discussion on the anisotropic ACS PSF that, for
the first time with a space-based weak lensing analysis, we have 
been able to characterise directly from our data without
having to assume long-term PSF stability.  Long-term PSF stability
has been tested with the GEMS data and shown
to be true for the ACS only above the $\sim 5\%$ level.  With previous 
space-based analyses it has often been 
assumed that temporal changes in the mean 
PSF pattern can be accommodated by a low order correction. 
Percent level accuracy in the ACS PSF correction
would however be difficult to achieve with this assumption as
this work has shown that the both the PSF strength and pattern exhibits 
temporal variation.
We have identified
PSF temporal variation on the level of a few percent finding consistent
behaviour 
between the F850LP imaging and F606W imaging, even though the F850LP PSF is
quite different from the F606W PSF.  We have tested the success of
our PSF correction by measuring the star-galaxy cross-correlation and the 
B-type shear correlator which were both found to be consistent with zero on
angular scales $\theta>0.2$ arcmin.
Our semi-time-dependent method for PSF modelling therefore adequately 
corrects for the varying PSF distortion when we consider weak lensing shear
correlations as a
function of relative galaxy position averaged over the whole GEMS mosaic.
When we measure 
the shear power spectrum however, a statistic which is dependent on galaxy
shear as a function of absolute
galaxy position, we find a significant detection of E-B cross-correlations
most likely revealing the impact of not producing a fully time-dependent PSF
correction model. 
It is currently unclear where the variation in the GEMS ACS PSF arises but
its presence, also seen by \citet{JeeACS} and Rhodes et. al. (in prep), 
suggests that future HST cosmic shear surveys should be preferentially
observed in sequence to minimise the impact of PSF instabilities.

We have measured the commonly used 
top-hat shear variance statistic, performing an E/B mode decomposition.
We find significant non-lensing B-mode distortions at angular
scales $\theta < 1.5$ arcmin in contrast to the E/B decomposition 
of the shear correlation function where the B-modes were found to be
consistent with zero 
at angular scales $\theta >0.2$ arcmin.  This shows that the top-hat shear
variance statistic becomes contaminated by very small scale non-lensing
distortions out to fairly high angular scales, strongly
biasing the final result.  Note that this effect is also
seen in \citet{vWb04}.  The shear correlation function
does not suffer from this contamination as the very small scale non-lensing
correlations are removed from the analysis and it is therefore this statistic
along with the shear power spectrum
that we favour for cosmological parameter estimation.  We urge
future cosmic shear studies to perform E/B mode decompositions to test for
non-lensing distortions and employ statistical analyses other than the
easily biased top-hat shear variance statistic.

\subsection{Comparison with other cosmic shear surveys}

\begin{figure} 
\centerline{\epsfig{file=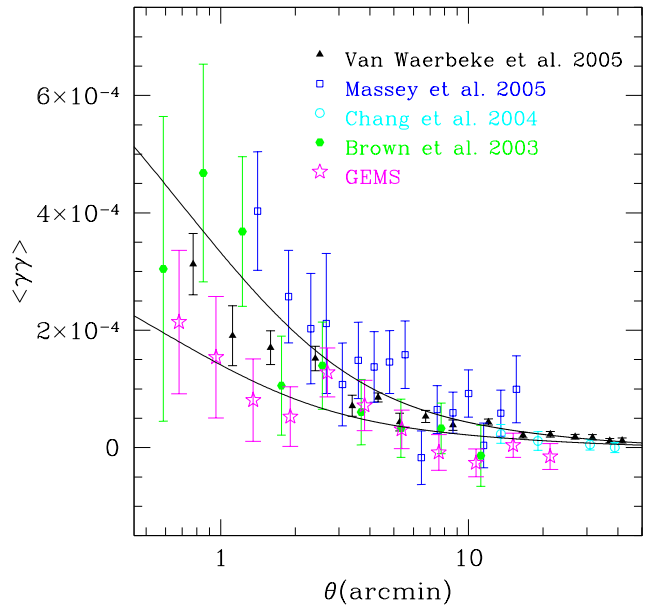,width=8.95cm,angle=0,clip=}} 
\caption{ Comparison of the total shear correlation function $E[\gamma
    \gamma]_\theta$ as measured from GEMS along with the most up-to-date
    shear correlation measurements from the other groups indicated.  
    Over-plotted are theoretical $\Lambda CDM$ models for a $z_m = 1$ survey
    with $\sigma_8 = 0.7$ (lower) and $\sigma_8 = 1.0$ (upper). Note all data
    points and errors have been scaled to a $z_m = 1$ survey
    using a $z_m^2$ redshift scaling.}
\label{fig:corcomp} 
\end{figure}

\begin{figure} 
\centerline{\epsfig{file=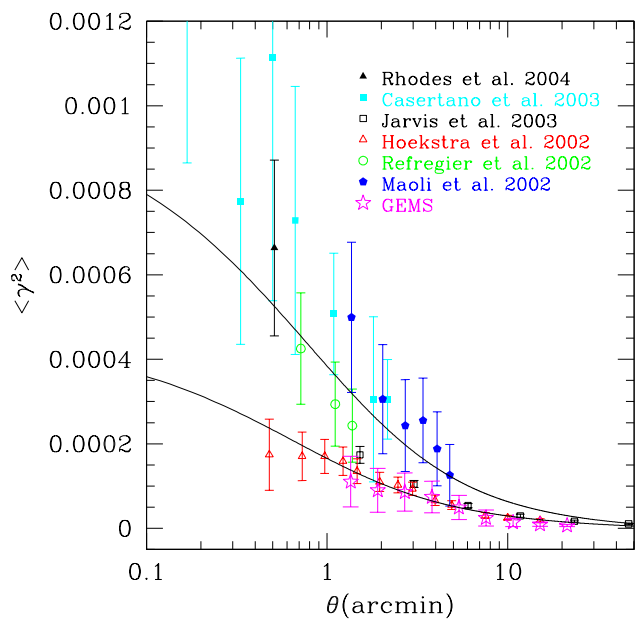,width=8.95cm,angle=0,clip=}} 
\caption{ Comparison of the top-hat shear variance $E[\gamma^2]_\theta$ as
    measured from GEMS along with the most up-to-date 
    top-hat shear variance measurements from the other groups indicated. 
    We show only the B-mode free GEMS top-hat shear variance results with
    $\theta \ge 1.5$ arcmin. 
    Over-plotted are theoretical $\Lambda CDM$ models for a $z_m = 1$ survey
    with $\sigma_8 = 0.7$ (lower) and $\sigma_8 = 1.0$ (upper). Note all data
    points and errors have been scaled to a $z_m = 1$ survey
    using a $z_m^{1.85}$ redshift scaling.}
\label{fig:tophatcomp} 
\end{figure}

Figures~\ref{fig:corcomp} and~\ref{fig:tophatcomp} compare the GEMS results
with the most up-to-date results from other cosmic shear surveys that
have placed constraints on $\sigma_8$ and $\Omega_m$.  The cosmic
shear signal scales with the depth of the survey and so we have
introduced a median redshift scaling\footnote{For the results from
  \citet{Chang} 
we scale assuming the median radio source redshift to be $z_m = 2.0$.  For the
results from \citet{Casertano} we convert the measurement from the top-hat
variance in square cells to the top-hat variance in circular cells using a
$1/\sqrt{\pi}$ scaling \citep{BRE} and then scale the results 
using the median redshift derived by \citet{RRG02} for the same data set.  
For the \citet{Jarvis} results we scale by $z_m = 0.6$ (private communication)
and for all other results, we use the quoted median redshift.}
of the data points and errors to
bring the different results in 
line with a survey of median redshift $z_m = 1.0$.  For the shear correlation
function (Figure~\ref{fig:corcomp}) we scale by $z_m^2$, as suggested by the
numerical simulations of \citet{Barber} and for the top-hat shear variance
(Figure~\ref{fig:tophatcomp}) we
scale by $z_m ^{1.85}$ \citep{RhodesSTIS}.  We preferentially show the shear
correlation function for 
surveys that have measured the top-hat shear variance in addition, due to our
concern with the use of the top-hat shear variance statistic.  We only show
the B-mode free  
GEMS top-hat shear variance results with $\theta \ge 1.5$ arcmin. These comparisons
show broad agreement between the shear correlation measurements and a poorer
agreement between the top-hat shear variance measurements.  As discussed in
Section~\ref{sec:shear_variance}, the top-hat shear variance appears to be
easily contaminated at large 
scales by small scale systematic errors and we propose that this contamination,
not always quantified, is at least a partial cause of the differing results.
Other possibilities are potential calibration biases arising from 
differences between the various shear measurement
methods (compare for example \citealt{erben} and \citealt{Baconsims}), differences
in the median redshift determination and sampling variance.

The results shown in Figure \ref{fig:corcomp} and Figure \ref{fig:tophatcomp}
yield measurements 
of $\sigma_8$ ranging from $\sigma_8 \simeq 0.7$ to $\sigma_8 \simeq 1.1$ for
a value of 
$\Omega_m= 0.3$.  This can be compared with results from the WMAP CMB
experiment \citep{WMAP} that finds $\sigma_8 = 0.9 \pm 0.1$ from the WMAP data
alone and $\sigma_8 = 0.84 \pm 0.04$ when the WMAP data is combined with other
data sets.  Results
from cluster abundance measurements range from $\sigma_8 = 0.7$ to
$\sigma_8 = 1.0$ (see \citealt{Pierpaoli03}, who find 
$\sigma_8 = 0.77 \pm 0.05$, and references therein).
Our measurement is at the lower end of all these results which we may
expect in light of the fact that the CDFS is a factor of two under-dense in
massive galaxies \citep{Wolf03}.  If we assume that massive galaxies trace
the underlying dark matter distribution, then we would 
expect a low measurement of $\sigma_8$ from this field when compared to
the global $\sigma_8$ value.  Combining
GEMS data with other wide-field space-based mosaics, such as the COSMOS
survey\footnote{{\it www.astro.caltech.edu/$\sim$cosmos}} and the ACS pure 
parallel
survey, will reduce the effects of sampling variance in order to obtain a good
estimate of the Universal value of $\sigma_8$ from HST.

\section{Acknowledgements}
This work is based on observations taken with the NASA/ESA Hubble Space
Telescope, which is 
operated by the Association of Universities for Research in Astronomy,
Inc. (AURA) under NASA contract NAS5-26555.  Support for the GEMS project was
provided by NASA through grant number GO-9500 from the Space Telescope Science
Institute, which is operated by AURA for NASA, under contract NAS5-26555. CH
and HWR acknowledge financial support from GIF.  MLB
and CW were supported by PPARC fellowships.  
EFB and SFS acknowledge financial support provided through the European
Community's Human Potential Program under contract HPRN-CT-2002-00316,
SISCO (EFB) and HPRN-CT-2002-00305, Euro3D RTN (SFS).  SJ and DHM acknowledge
support from NASA under LTSA Grant NAG5-13063(SJ) and under LTSA Grant
NAG5-13102(DHM), issued through the Office of Space Science. 
We thank Justin Albert, Colin Cox, Richard Ellis, Yannick Mellier, John
Peacock, Jason Rhodes, 
Tim Schrabback, Ludovic Van Waerbeke and Martin White 
for many useful discussions, and 
also the referee for his/her helpful comments.

\bibliographystyle{mn2e}
\bibliography{ceh_2004}
\label{lastpage}

\end{document}